\documentclass[usegraphicx,usenatbib,useAMS]{mn2e}

\usepackage{times,amssymb}
\usepackage[english]{babel}
\usepackage[]{graphicx}
\usepackage[T1]{fontenc}
\usepackage[varg]{txfonts}
\usepackage{natbib}
\bibpunct{(}{)}{;}{a}{}{,}

\usepackage{aasmacros}

\title[Properties of the UV flux of SN~Ia]{Properties of the ultraviolet flux of
       type~Ia supernovae: an analysis with synthetic spectra of
       SN~2001ep and SN~2001eh}

\author[D.~N.~Sauer et al.]{D.~N.~Sauer,$^{1,5}$\thanks{email: {\tt dsauer@mpa-garching.mpg.de}} %
        P.~A.~Mazzali,$^{1,2,5}$
        S.~Blondin,$^{3,4,5}$
        M.~Stehle,$^{1}$
        S.~Benetti,$^{2}$\newauthor
        P.~Challis,$^{4}$
        A.~V.~Filippenko,$^{6}$
        R.~P.~Kirshner,$^{4,5}$
        W.~Li,$^{6}$ and
        T.~Matheson$^{7}$
        \\
$^{1}$Max-Planck-Institut f\"ur Astrophysik, Karl-Schwarzschild-Str. 1, 85748 Garching, Germany \\
$^{2}$Istituto Nazionale di Astrofisica-OAPd, vicolo dell'Osservatorio 5, 35122 Padova, Italy\\
$^{3}$European Southern Observatory, Karl-Schwarzschild-Str. 2, 85748 Garching, Germany\\
$^{4}$Harvard-Smithsonian Center for Astrophysics, 60 Garden Street, Cambridge, MA 02138, USA\\
$^{5}$Kavli Institute for Theoretical Physics, University of California, Santa Barbara, CA 93106-4030, USA\\
$^{6}$Department of Astronomy, University of California, Berkeley, CA 94720-3411, USA\\
$^{7}$National Optical Astronomy Observatory, 950 N. Cherry Avenue, Tucson, AZ 85719-4933, USA
}

\date{\today}
\newcommand{\snia}{SN~Ia}
\newcommand{\sneia}{SNe~Ia}
\newcommand{\tsim}{\ensuremath{\sim}}
\newcommand{\lam}{\ensuremath{\lambda}}
\newcommand{\nifs}{\ensuremath{^{56}\rm{Ni}}}
\newcommand{\cofs}{\ensuremath{^{56}\rm{Co}}}
\newcommand{\fefs}{\ensuremath{^{56}\rm{Fe}}}

\newcommand{\kms}{\ensuremath{\rm{km\,s}^{-1}}}
\newcommand{\ergs}{\ensuremath{\rm{erg\,s}^{-1}}}
\newcommand{\bv}{\ensuremath{B\!-\!V}}

\newcommand{\ebv}{\ensuremath{E(\bv)}}

\begin{document}

\maketitle

\begin{abstract}
  The spectral properties of type Ia supernovae in the ultraviolet (UV) are
  investigated using the early-time spectra of SN~2001ep and SN~2001eh obtained
  using the {\em Hubble Space Telescope (HST)}.  A series of spectral models is
  computed with a Monte Carlo spectral synthesis code, and the dependence of
  the UV flux on the elemental abundances and the density gradient in the outer
  layers of the ejecta is tested.  A large fraction of the UV flux is formed by
  reverse fluorescence scattering of photons from red to blue wavelengths. This
  process, combined with ionization shifts due to enhanced line blocking, can
  lead to a stronger UV flux as the iron-group abundance in the outer layers is
  increased, contrary to previous claims.
\end{abstract}


\section{Introduction}
\label{sec:intro}

Type Ia supernovae ({\sneia}) are among the most important tools to study the
expansion history of the Universe and to set constraints on the properties of
dark energy
(\citealt{riess98,perlmutter99,tonry03,riess04,astier06,riess07,wood-vasey07};
see \citealt{leibundgut01,filippenko04,filippenko05} for reviews). With the
empirical calibration of their peak luminosity
\citep{phillips93,phillips99,jha07}, it is feasible to probe the properties of
the Universe out to a redshift of $z \approx 1$ and higher
\citep{goldhaber01,tonry03,riess04,riess07}.  Understanding the formation of
the ultraviolet (UV) spectrum of SNe Ia is important for two reasons: the
optical spectrum is strongly affected by line blocking in the UV, and optical
observations of high-redshift supernovae primarily cover the rest-frame UV
bands.  Comparisons between local and distant SNe Ia have recently identified
the UV as an important region to search for possible effects of evolution
\citep{foley07a,bronder08,ellis08}.

Efforts have been made to increase the sample of local {\sneia} observed in the
UV by space-borne telescopes.  The amount of data available for {\sneia} in the
UV is very small and mostly limited to single epochs -- usually around (or well
after) maximum light \citep{panagia03,foley08}.  Prominent examples include the
{\it Hubble Space Telescope (HST)} spectra of SN~1992A ($5\,$d and $45\,$d
after $B$-band maximum, \citealt{kirshner93}) and the very early {\it
International Ultraviolet Explorer (IUE)} spectrum of SN~1990N ($14\,$d before
$B$-band maximum, \citealt{leibundgut91}). Recently, the ultraviolet/optical
telescope (UVOT) onboard the {\it Swift} satellite \citep{gehrels04} has also
provided valuable UV data on {\sneia} \citep{immler06}.

To interpret such data correctly, it is necessary to understand the formation
of the UV flux in {\sneia} and its sensitivity to physical parameters of the
explosion and ambient medium.  Several groups have studied the properties of
the UV flux using radiative transfer models of differing complexity
\citep{pauldrach96,fisher97,hoeflich98,mazzali00,lentz01}.

In this paper, we use radiative transfer models to study the properties of
SN~2001eh and SN~2001ep, which were observed as part of the Supernova INtensive
Study (SINS; GO--9114; PI R. P. Kirshner) with {\it HST}.  Based on the
spectral models, we investigate how a change of physical parameters may affect
the observed UV flux of a {\snia}. In particular, when using {\sneia} as
standard candles to measure cosmological parameters, it is important to know if
parameters that are likely to change over a large range of redshifts can
introduce systematic errors in the distance measurement. Among the parameters
that may be affected by the environment of {\snia} progenitors is the abundance
of heavy elements in the outer shells of the ejecta. These layers are of
particular importance for the formation of the UV spectrum during the
photospheric phase.  The two supernovae chosen for this study are at the
opposite ends of the luminosity distribution of normal {\sneia}.

From spectral analysis of {\sneia}, we know that the composition is dominated
by a mixture of intermediate-mass elements, such as Si, S, and Mg, and
iron-group elements (Ni, Co, and Fe). At early epochs, until a few weeks after
maximum brightness, the spectrum of a {\snia} consists of a number of broad
absorption troughs that generally originate from blends of various lines. (For
a review of the observed properties of {\sneia}, see \citealt{filippenko97} and
\citealt{leibundgut00}.)  The emission of {\sneia} is entirely powered by the
radioactive decay of {\nifs} created in the explosion. The $\gamma$-ray photons
from radioactive decay undergo Compton scattering, creating fast electrons
which deposit their energy via collisions in the expanding ejecta
\citep{truran67,colgate69}.  The non-thermal continuum seen at early epochs
around maximum light is formed by the overlap of thousands of weaker lines.
The spectrum peaks at a wavelength between $4000$ and $5000\,${\AA} and drops
steeply in the UV, bluewards of the Ca H\&K lines, which typically have an
absorption minimum at about $3750\,${\AA} near maximum light \citep{blondin06}.

The UV wavelength bands of {\sneia} comprise only a small fraction of the total
emitted light, and are generally characterized by the absence of strong
features.  Singly and doubly ionized species of heavy elements such as Ti, Cr,
Fe, Co, and Ni have a large number of line transitions in the UV bands that
effectively block most of the emission in these bands
\citep{karp77,hoeflich93,pauldrach96}.  \citet{lucy99} and \citet{mazzali00}
find that essentially all UV photons created at deeper layers in the ejecta are
immediately absorbed and down-scattered to lower energies (longer wavelengths)
where the line opacity is smaller.  The crucial mechanism for the emission of
photons in the UV is found to be a reverse-fluorescence process where
absorption of red photons is followed by emission in the blue and UV bands
\citep{lucy99,mazzali00}.  These processes have a significantly lower
probability of occurrence but in the outer region of the ejecta the emission of
bluer photons via reverse fluorescence can be significant because there are
many more red photons available than blue ones such that the lines in the blue
are not saturated. Blue photons can escape from a line rather than getting
re-absorbed locally. The UV spectrum is predominantly shaped by this
fluorescence spectrum originating from a confined layer of high-velocity
material, which is small compared to the dimensions of the expanding ejecta
\citep{mazzali00}. The shape of the emission spectrum is determined by the
atomic properties of the ions in the emission layer.  This makes the UV an
interesting wavelength band in which to study possible progenitor scenarios
because the progenitor star may leave a characteristic imprint on the
composition of the outermost layers of the expanding shell.

\citet{lentz01} investigated the dependence of the UV flux of {\sneia} on the
composition of the outermost layers of the ejecta, adopting different fractions
of solar metallicity for the material above $v \approx 14\,000\,${\kms}. They
found that overall the UV flux decreases with increasing metallicity. They also
pointed out, however, that the larger number of strong absorption lines with
higher metallicity can cause a back reaction of the blocked radiation field on
the temperature and ionization structure (line blanketing), which is difficult
to disentangle from the line-blocking effect.  Therefore, the UV flux may vary
non-linearly with a change of composition.

In this paper we investigate the sensitivity of the UV spectrum to
modifications of the composition in the outer layers. We also consider
variations in the density structure of the underlying explosion model.

The paper is organized as follows.  In Section~\ref{sec:obs} we present
observations of SN~2001ep and SN~2001eh covering the UV wavelength range as an
example of {\sneia} with different UV properties.  Section~\ref{sec:models}
outlines the radiative transfer code used for this work and an analysis of the
observed spectra with the help of synthetic spectra.  The radiative transfer
models for both supernovae are used as a basis for a series of models to study
the properties of the UV flux with respect to changes in composition and
density structure. The results of this series and the physical processes
responsible for the formation of the UV flux are discussed in
Section~\ref{sec:series}.  Section~\ref{sec:conclusions} gives the conclusions
of our study.

\section{Observations}
\label{sec:obs}

We present observations and synthetic spectra of SN~2001eh and SN~2001ep as
examples of SN~Ia that exhibit significantly different spectral properties,
especially in the UV. A more detailed discussion of the complete data available
for the two supernovae ($UBVRI$ light curves, optical spectra, and one
additional UV spectrum for each SN) will be presented in a forthcoming paper
(Blondin et al., in preparation). Here we present the data relevant to the
study of the UV spectra discussed in this paper. The observations are
summarized in Table~\ref{obsdata}.

SN~2001eh was discovered on 9 Sep. 2001 (UT dates are used throughout this
paper) by M. Armstrong \citep{hurst01} and independently by M. Ganeshalingam
and W. D. Li \citep{ganeshalingam01} with the Katzman Automatic Imaging
Telescope (KAIT; \citealt{filippenko01}) as part of the Lick Observatory and
Tenagra Observatory Supernova Search (LOTOSS). It occurred in UGC 1162, which
have a NED\footnote{The NASA/IPAC Extragalactic Database (NED) is available at
http://nedwww.ipac.caltech.edu.} recession velocity of $cz=11131$\,{\kms} (from
The Updated Zwicky Catalog\footnote{{\tt http://tdc-www.harvard.edu/uzc/}};
\citealt{falco99}). A CCD spectrum obtained on 11 Sep. with the Shane 3-m
reflector at Lick Observatory revealed that SN 2001eh was a SN~Ia about a week
before maximum brightness \citep{ganeshalingam01}.

SN~2001ep was discovered on 3 Oct. 2001 by \citet{hutchings01} as part of
LOTOSS. It occurred in  NGC 1699, which has a NED recession velocity of
$cz=3901$\,{\kms} \citep{devaucouleurs91}. A CCD spectrum obtained on 10 Oct.
with the Fred L.  Whipple Observatory (FLWO) 1.5-m Tillinghast telescope and
FAST spectrograph \citep{fabricant98} showed that it was a SN~Ia near maximum
brightness \citep{matheson01}.

SN~2001eh and SN~2001ep were observed with $HST$ (Cycle 10) as targets of
opportunity. The data were obtained with the Space Telescope Imaging
Spectrograph (STIS) using a 52\arcsec$\times$0.2\arcsec slit. For both
supernovae we used the G430L grating to obtain ``optical'' spectra
(2900-5700\,{\AA}), the G230L grating to obtain near-UV spectra
(1600-3100\,{\AA}), and the G140L grating to obtain far-UV spectra
(1150-1700\,{\AA}). No clear signal was detected for any observation taken with
the latter setup. The reductions followed the standard STIS reduction pipeline.
Spectra were extracted from the processed images using the IRAF\footnote{IRAF
is distributed by the National Optical Astronomy Observatories, operated by the
Association of Universities for Research in Astronomy, Inc., under contract to
the National Science Foundation of the United States.} SPECRED long-slit
reduction packet.

The optical spectra presented in this paper were obtained with the FLWO 1.5-m
FAST spectrograph using a slit of width 3{\arcsec}. For both supernovae we used
a 300 line mm$^{-1}$ grating with two tilts: one covering \tsim$3700-7500\,${\AA},
the other \tsim$5200--9000\,${\AA}. The spectra were reduced and calibrated
employing standard techniques in IRAF and our own IDL routines for flux
calibration.

\begin{table*}
\caption{UV and optical spectra of SN~2001eh and SN~2001ep}
\label{obsdata}
\centering
\begin{tabular}{ccccccc}
\hline\hline
IAU name & UT Date & Telescope+ & Slit      & Grating & Range & Exptime \\
         &         & Instrument & (\arcsec) &         & (\AA) & (s)   \\
\hline
2001eh & 2001-09-25.67 & $HST$+STIS     & 52$\times$0.2 & G140L & 1140-1730 & 1340          \\
2001eh & 2001-09-25.76 & $HST$+STIS     & 52$\times$0.2 & G230L & 1570-3180 & 2$\times$2700 \\
2001eh & 2001-09-25.66 & $HST$+STIS     & 52$\times$0.2 & G430L & 2900-5700 & 500           \\
2001eh & 2001-09-26.35 & FLWO 1.5m+FAST & 3             & 300   & 3720-7540 & 1200          \\
2001eh & 2001-09-26.37 & FLWO 1.5m+FAST & 3             & 300   & 5200-9040 & 1200          \\
\hline
2001ep & 2001-10-27.99 & $HST$+STIS     & 52$\times$0.2 & G140L & 1140-1730 & 1630,2690     \\
2001ep & 2001-10-28.09 & $HST$+STIS     & 52$\times$0.2 & G230L & 1570-3180 & 2690          \\
2001ep & 2001-10-27.95 & $HST$+STIS     & 52$\times$0.2 & G430L & 2900-5700 & 200           \\
2001ep & 2001-10-25.42 & FLWO 1.5m+FAST & 3             & 300   & 3720-7540 & 1200          \\
2001ep & 2001-10-25.44 & FLWO 1.5m+FAST & 3             & 300   & 5200-9040 & 1200          \\
\hline
\end{tabular}
\end{table*}

\begin{figure}
  \centering
  \includegraphics[width=8.5cm]{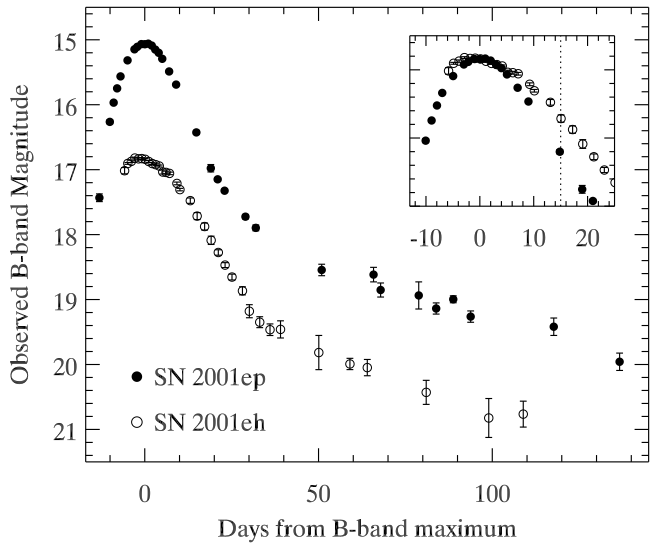}
  \caption{KAIT $B$-band light curves of SN~2001eh and SN~2001ep. The
    inset shows both light curves out to \tsim$20\,$d past $B$-band
    maximum, normalized in magnitude at the peak. The dotted line
    at +15\,d highlights the difference in $\Delta m_{15}(B)$ between
    the two supernovae (see text for details).}
  \label{fig:lcfig}
\end{figure}

The KAIT $B$-band light curves for SN~2001eh and SN~2001ep are shown in
Fig.~\ref{fig:lcfig}. Both are sufficiently well sampled to enable a direct
measurement of the decline-rate parameter $\Delta m_{15}(B)$. For SN~2001eh we
find $\Delta m_{15}(B)=0.71\pm0.05$, while for SN~2001ep we find
$\Delta m_{15}(B)=1.41\pm0.04$. Thus, SN~2001eh is a ``slow decliner,'' while
SN~2001ep is at the fast end of typical ``intermediate-decliner'' SNe~Ia.

The differences in light-curve shape between the two supernovae are reflected
in their spectra. Using the Supernova Identification (SNID) cross-correlation
code of \citet{blondin07}, we find that the best-match for SN~2001eh is the
overluminous SN~1999aa ($\Delta m_{15}(B)=0.75\pm0.02$;
\citealt{krisciunas00}), and that the best-match template spectrum for
SN~2001ep is the ``transitional'' SN~2004eo ($\Delta m_{15}(B)=1.45\pm0.04$;
\citealt{pastorello07}).

We fitted the complete $UBVRI$ photometric data with the MLCS2k2 light-curve
fitter of \citet{jha07}. The time of $B$-band maximum is found to be 2001 Sep.
17.53 (JD $=2452170.03\pm0.71$) for SN~2001eh and Oct. 17.72 (JD
$=2452200.22\pm0.45$) for SN~2001ep. The derived line-of-sight extinction in
the host galaxy is $A_V=0.08\pm0.06$ and $A_V=0.43\pm0.12$ mag, respectively.
The Milky Way foreground reddening is small for both supernovae ($E(B-V)=0.06$\,mag
for SN~2001eh and $E(B-V)=0.05$\,mag for SN~2001ep; \citealt{schlegel98}).

The UV and optical spectra of each SN were combined into a single spectrum.
The far-UV data taken with the G140L grating were not included in either of the
combined spectra given the lack of signal. The mean UT date for the SN~2001eh
spectrum is 2001-09-25.93, corresponding to \tsim$9\,$d past $B$-band maximum.
The mean UT date for the SN~2001ep spectrum is 2001-10-26.88, also
corresponding to \tsim$9\,$d past $B$-band maximum. Assuming a $B$-band rise
time of \tsim$19.5\,$d \citep{conley06, riess99b}, both spectra correspond to
28-29\,d past explosion. In what follows we assume that both spectra are at
$t=29$\,d past explosion.

\section{Spectral models}
\label{sec:models}

\subsection{The radiative transfer model}
\label{sec:models:method}

To model synthetic spectra of {\sneia} at early epochs, we use a Monte Carlo
spectral synthesis code with an approximate description of non-LTE (i.e., not
assuming local thermodynamic equilibrium, LTE). Here we outline the most
important aspects of our method as they are relevant for the discussion in this
work.  More detailed descriptions can be found in \citet{abbott85},
\citet{mazzali93b}, and \citet{lucy99}, as well as in \citet{mazzali00}. This
last reference also gives the sources of atomic data used by the code.

For the underlying explosion model we take the density structure from the
one-dimensional deflagration model W7 \citep{nomoto84}. The composition is
adjusted to match the observed spectrum.  For this study we use the version of
the code that allows a radial variation of the composition within the ejecta
described by \citet{stehle05}.

The code is based on the Schuster-Schwarzschild approximation -- that is, we
assume that the radiative energy is entirely emitted at the inner boundary of
the computational grid with the spectral shape of a blackbody. The spectra are
calculated assuming  stationarity and ignoring any time-dependences.  This is a
good approximation at least for 1D  models of SN~Ia later than a few days after
explosion \citep{kasen06}.  We do not consider the deposition of energy by
$\gamma$-ray photons from the {\nifs} and {\cofs} decays within the atmosphere,
nor do we try to derive the total luminosity directly from the {\nifs} mass.
Therefore, luminosity is a free parameter. To constrain the model, the location
of the inner boundary (``photosphere''\footnote{We generally avoid the term
``photosphere'' for this inner boundary. In contrast to stars, where this term
refers to the radius at which the atmosphere becomes optically thick with
respect to a wavelength-independent {\em mean} optical depth, this concept
cannot be used in hydrogen-deficient supernovae, where the optical-depth scale
is strongly wavelength-dependent owing to the shallow density gradient and the
absence of strong continuum opacities \citep[see, e.g.,][]{sauer06a}.}) has to
be specified in terms of a velocity $v_{\rm ib}$ and the elapsed time $t$ since
the explosion, assuming homologous expansion ($r(t)=vt$) of the ejecta.  In
addition to those physical input parameters, observational parameters such as
distance and reddening have to be specified to allow direct comparison of the
synthetic and the observed spectrum.

The code uses an approximate description of non-LTE to derive the atomic level
populations and the ionization equilibrium.  Collisional and continuum
processes are not considered, apart from electron scattering. Since the opacity
in {\snia} ejecta is largely dominated by the opacity of Doppler-shifted and
broadened spectral lines, the true continuum plays only a minor role. Given the
low densities in the ejecta, collisional processes are also not expected to
have a major impact on the state of the gas. It is not possible to derive a
consistent gas temperature in this approximation.  Thus, following
\citet{mazzali93b}, the local gas temperature is approximated to be $90\%$ of
the radiation temperature $T_{\rm R}$ of the radiation field, which is
determined from the mean energy $\bar{x}$ of a photon in a Planckian radiation
field,
\begin{equation}
  \bar{x} = \frac{h\bar{\nu}}{k_{\rm B}T} = \frac{h}{k_{\rm B}T}\frac{\int_{0}^{\infty}
  \nu B_{\nu}\,d\nu}{\int_{0}^{\infty}
  B_{\nu}\,d\nu} = \frac{360}{\pi^{4}}\zeta(5)\approx3.832
  \label{eq:temperature}
\end{equation}
(see also \citealt{lucy99a}). The temperature of the blackbody at the inner
boundary is determined in an iteration cycle, using the constraint that the
desired luminosity $L$ is obtained at the outer boundary. Iteration is
necessary because the amount of radiation that is scattered back into the core
is not known {\it a priori}.

A crucial quantity for the derivation of the ionization and excitation
equilibrium is the ``equivalent dilution factor'' $W$ introduced by
\citet{mazzali93b}. This factor, which is derived from the nebular
approximation for the radiation field in an optically thin medium, is used to
modify the LTE Saha-Boltzmann excitation formula. While in a gaseous nebula $W$
describes a purely geometrical dilution of the radiation field, the procedure
developed for supernova ejecta derives $W$ directly from the assumed
relationship between the Planck function at the radiation temperature
$T_{\rm R}$ and the radiation field $J$:
\begin{equation}
  J=WB(T)=W \frac{\sigma}{\pi}T^{4}_{\rm R},
  \label{eq:w-def}
\end{equation}
where $\sigma$ denotes the Stefan-Boltzmann constant. This approximation has
been shown to produce a reasonably good representation of non-LTE ionization
and excitation in supernova ejecta \citep{mazzali93b}.

Line processes are treated using the line-branching formalism introduced by
\citet{lucy99} and \citet{mazzali00}. While earlier work with this code assumed
pure resonant scattering of photons in spectral-line transitions, this
improvement allows photons to be re-emitted in a different transition than the
one by which they were absorbed. \citet{mazzali00} showed that the inclusion of
branching processes is necessary to reproduce the flux in the UV.

Following the description of \citet{lucy99}, the emergent spectrum is derived
by a formal solution of the transfer equation using line and continuum source
functions that are extracted from the Monte Carlo simulation. The advantage of
this method over using emerging Monte Carlo packets directly is a significantly
reduced number of Monte Carlo packets to obtain a converged spectrum with low
Monte Carlo noise. Because the UV flux of a {\snia} is intrinsically very low,
one must nevertheless ensure that the number of packets used in the Monte Carlo
experiment is sufficient to provide a converged solution for the source
functions. We performed convergence tests in the UV and determined that the
relative scatter of the spectrum caused by Monte Carlo noise in the wavelength
range between $1500$ and $3000\,${\AA} is generally well below $10$\% for
$500\,000$ photon packets.  With this number the scatter in the optical
spectrum is less than $0.4$\%.

\subsection{Synthetic spectra for SN~2001ep and SN~2001eh}
\label{sec:models:models}

\begin{figure}
  \begin{center}
    \includegraphics[width=8.3cm]{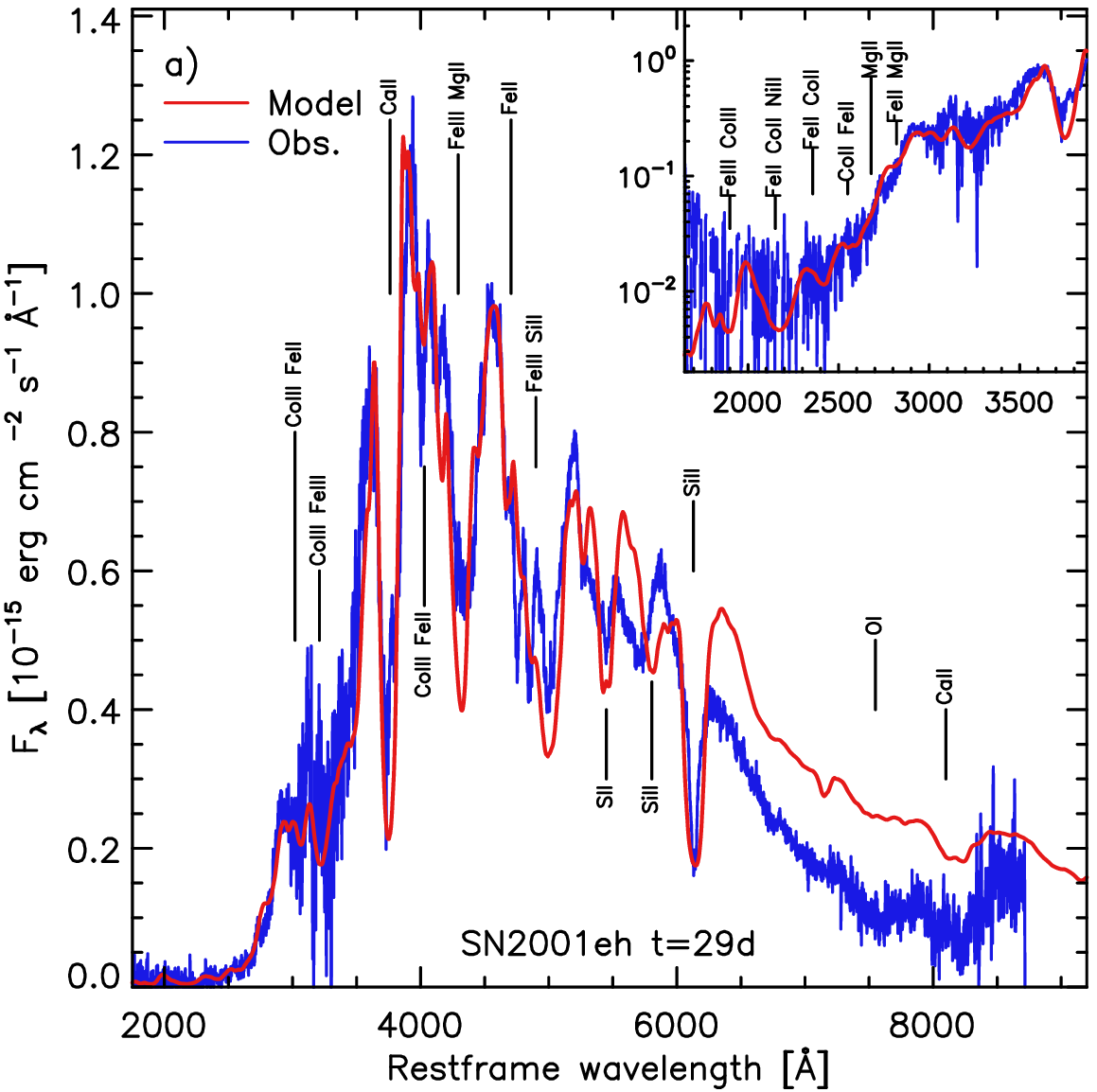}
    \includegraphics[width=8.3cm]{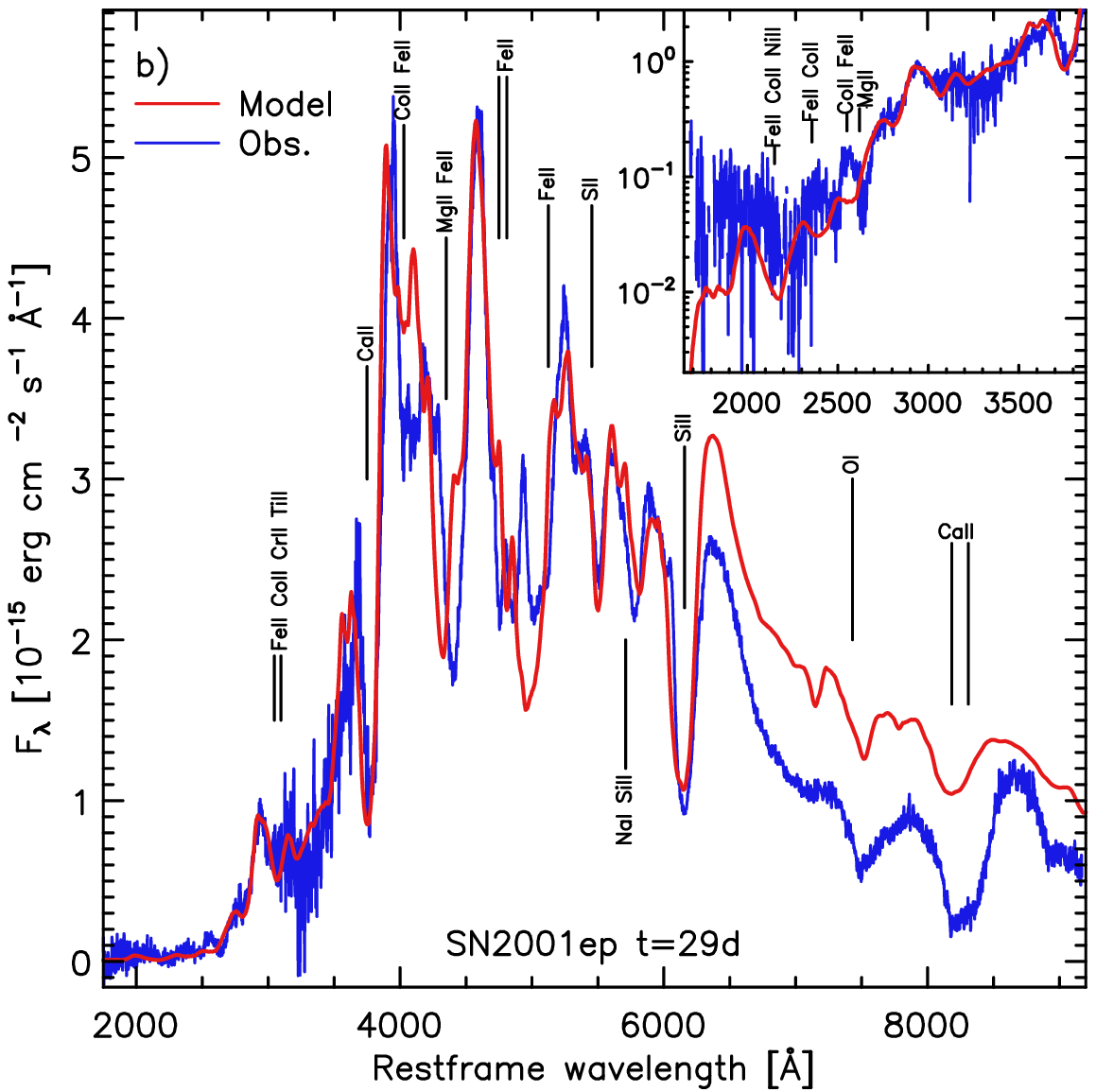}
  \end{center}
  \caption{Synthetic and observed spectra of (a) SN~2001eh and (b) SN~2001ep,
  both at $t=29\,$d after explosion. The inserts show the UV part of the
  respective spectra with a logarithmic scale.  The labels indicate the ions
  that dominate the formation of the various absorption features. The blue and
  UV parts of the spectrum are formed by a blend of a large number of lines,
  precluding a complete identification of the contributors.  } \label{fig:models}
\end{figure}

The models and the observed spectrum of SN~2001eh are shown in
Fig.~\ref{fig:models}a; those of SN~2001ep are shown in Fig.~\ref{fig:models}b.
The inserts in the figures show the UV part of the spectra on a logarithmic
scale.  The direct comparison of both observed spectra clearly shows that these
supernovae exhibit fairly different spectral appearances. In SN~2001eh the UV
flux blue of \ion{Ca}{2} H\&K at \tsim$3750\,${\AA} is stronger with relation
to the optical spectrum.  The characteristic line features are fairly narrow
and are found at lower velocities than in the spectrum of SN~2001ep.

Modelling the spectra requires knowledge of the distance and reddening of the
supernovae. For the distance we use the data available for the respective host
galaxies listed in NED. The redshift of the host galaxy of SN~2001eh (UGC~1162,
$z=0.037$) corresponds to a distance modulus of $m-M=35.9\,$mag assuming
$H_{0}=73\,$km\,s$^{-1}$\,Mpc$^{-1}$  \citep{riess05}. NGC~1699, the host
galaxy of SN~2001ep, has a redshift of $z=0.013$, resulting in a distance
modulus of $m-M=33.6\,$mag.  For the model of SN~2001eh we assume a total
reddening of $\ebv=0.07$ mag, while for SN~2001ep we use $\ebv=0.15\,$mag. The
value for the reddening of SN~2001ep is at the lower end of the uncertainty
range estimated from observational indicators (see Section~\ref{sec:obs});
higher values for {\ebv}, however, require a higher luminosity and lead to a
shift in the ionization towards doubly ionized species of Fe and Co that are
not seen in the observed spectrum.

For both supernovae the epoch was taken to be $29\,$d after explosion (see
Section~\ref{sec:obs}).  The synthetic spectrum of SN~2001eh was modelled with
a luminosity of $L=1.15\times10^{43}\,${\ergs}. For the spectrum of SN~2001ep
we need a lower luminosity of  $L=7.98\times10^{42}\,${\ergs}.   For both
models we adopted the density structure from the W7 explosion model
\citep{nomoto84}; the composition was adjusted to fit the observed spectrum.
To model SN~2001eh we set the velocity of the inner boundary to
$v_{\rm ib}=5200\,${\kms}.  For SN~2001ep we used $v_{\rm ib}=5000\,${\kms}.
Above the inner boundary, the ejecta were divided into several radial zones
where the composition was adjusted independently. The composition of the outer
zones was determined using the spectra of earlier epochs available for both
supernovae. The outcome of theoretical explosion models such as W7 and the
results of tomography studies of SN~Ia with similar properties (SN~2002bo
\citep{stehle05} for SN~2001eh and SN~2004eo \citep{mazzali08} for SN~2001ep)
served as a rough guideline. The inner region ($v\lesssim8000\,${\kms}) of the
ejecta of SN~2001eh contain mostly ($>80\%$) elements generated by burning to
nuclear statistical equilibrium (NSE; Fe-group material, mostly {\nifs}).  This
is followed by a layer containing \tsim$60\%$ intermediate mass elements (IME)
(Si, S, Mg, Ca) and \tsim$20\%$ of NSE material out to $12\,000\,${\kms}. The
layers above $12\,000\,${\kms} are dominated by unburned O and C ($58$-$98\%$)
with small amounts of IME (between $2$ and $20$\%) and NSE material ($0.5$
-$4.5\%$), decreasing outwards. In the less luminous SN~2001ep the IME-rich
region extends to lower velocities (\tsim$6000\,${\kms}) and the O-rich layer
reaches down to \tsim $9000\,${\kms}.  The mass fraction of NSE material
appears to drop rapidly ($\lesssim25\%$) above \tsim$7000\,${\kms}. In the
outermost layers above \tsim $10\,000\,${\kms} the mass fraction of  IME
material drops from \tsim $40$ to $1\%$ outwards, NSE material from $3$ to
$0.2\%$; The NSE abundance in the outer layers is somewhat lower than in
SN~2001eh which is in line with the findings of \citet{mazzali07}.  The
presence of small amounts of NSE material in the outermost layers is, however,
necessary to fit the UV part of the spectrum in both supernovae. The original
W7 density structure only extends to a velocity of $24\,000\,${\kms}.  To
ensure that the choice of the maximum velocity of the computational grid does
not affect the outcome of the model, the density structure is extrapolated to
higher velocities using the power-law index of the outer zones of the
input-density structure. In the model series discussed later in this paper, the
maximum velocities is set to large values (\tsim$70\,000\,${\kms}) to avoid
artefacts generated by this boundary.

The models fit the spectra fairly well in the UV and optical parts of the
spectrum.  The offset in the wavebands redwards of the \ion{Si}{2} feature at
\tsim$6100\,${\AA} is caused by the breakdown of our assumption of a thermal
lower boundary. The opacity at these wavelengths is dominated by electron
scattering, which does not modify the spectral properties of the radiation
field. The radiation at these wavelengths originates from layers below our
adopted inner boundary and is shaped by a large number of individually weak
lines \citep{kasen06a}. With our assumption of stationarity and the chosen set
of atomic data, the shape of the pseudo-continuum cannot be correctly described
by the model even if extended to deeper layers \citep{sauer06a}.  Additionally,
a description of the ejecta below the lower boundary implies that we rely on
specific explosion models, which introduce additional uncertainties.  The
absorption features blue of \tsim$2700\,${\AA} in the spectrum of SN~2001ep
(insert of Fig.~\ref{fig:models}b) appear to be too blue in the synthetic fit.
This mismatch probably indicates that the outer part of the density structure
used to compute the model is not appropriate. This is not surprising given that
the W7 model generated more {\nifs} than what is expected for SN~2001ep and may
therefore also have a larger kinetic energy than required to describe this SN.
Another indication of an incorrect density structure is that the model fails to
reproduce the separated \ion{Fe}{2} features around $5000\,${\AA} which produce
the apparent emission feature at \tsim$4950\,${\AA}.

The major absorption features are labelled in Fig.~\ref{fig:models}.  However,
especially towards the blue and UV, where the features are blends of many
different lines, ions other than those indicated in the plot also contribute to
the shape of the absorption.  SN~2001eh is bluer than SN~2001ep and has a
different spectral appearance at the same epoch. The spectrum of SN~2001eh
contains a significant contribution from doubly ionized species such as
\ion{Fe}{3} and \ion{Co}{3}, while the spectrum of SN~2001ep is dominated by
lines of singly ionized species.   The characteristic \ion{Fe}{2} features
(e.g., between $4500\,${\AA} and $5000\,${\AA}) are significantly weaker in
SN~2001eh than in SN~2001ep.  The comparison of the dominant ionization stages
indicates that the gas temperatures in the ejecta of SN~2001eh are higher than
in SN~2001ep. In the models of SN~2001ep  the temperatures at similar
velocities are generally $500$-$1000\,$K lower than in the SN~2001eh model.

More differences are visible in the region bluewards of \ion{Ca}{2} H\&K.
Relative to the optical emission peaks, SN~2001eh has substantially more flux
in this region than SN~2001ep.  Further to the UV the \ion{Mg}{2} $2800\,${\AA}
absorption, which dominates the shape of the spectrum in the region around
$2600\,${\AA}, is stronger in SN~2001ep.

The differences seen in the UV result from different compositions or densities
of the outermost layers of the ejecta. In particular, Fe-group ions from Ti to
Ni have a large number of spectral lines in the UV that lead to large optical
depths out to the highest velocities where this part of the spectrum is shaped.

\section{The UV flux in {\snia} --- a model series}
\label{sec:series}

using a series of parameterized models where we modified first the composition
and then the density gradient in the outer layers.  Composition is expected to
be the primary quantity responsible for changes in the UV spectrum \citep[see
e.g.,][]{lentz01}. In addition, the imprint of a progenitor is likely to
manifest itself as a variation of the composition in the outer layers.  A
modified density structure can also be responsible for changes in the
appearance of the UV spectrum.  Different explosion models predict variations
in density, in particular at the higher velocities. Specifically, models with a
deflagration-to-detonation transition (DDT) occurring at some stage during the
explosion accelerate the matter to higher velocities resulting in higher
densities and more burned material in the outer layers of the ejecta
\citep{khokhlov91,hoeflich95,iwamoto99,roepke07,roepke07c}.

\subsection{Variation of the composition}

\subsubsection{Setup}

We use the models discussed in the previous section to set up a parameterized
model series in which the abundance of heavy elements is varied stepwise in the
outermost zone. For the first test series we keep the density structure of the
model unchanged and balance the modification of heavy-element abundances with a
change of the oxygen abundance. In the outer layers of the models, O is the
most abundant element and causes very few visible features in the early-time
spectrum of {\sneia}.  Therefore, even drastic changes in the oxygen abundance
do not affect the spectral appearance; the models primarily reflect the effects
of a variation in the abundance of heavy elements. (For the most extreme cases
a change of the electron density may become significant. A composition
dominated by oxygen can provide more electrons per unit mass than IME or
Fe-group elements.  Most of this effect is, however, counterbalanced by the
higher ionization potential of oxygen resulting in a larger fraction of neutral
atoms than for heavier elements.)

Within each series all model parameters are kept constant and only the
abundance of a specific element or group of elements is increased or decreased
above a velocity $v_{\rm shell}$. For the models of SN~2001eh we set
$v_{\rm shell}=15\,000\,${\kms}, while for the SN~2001ep-based models we use
$v_{\rm shell}=14\,500\,${\kms}.  The velocities are chosen to be roughly the
range at which the radiation field in the optical wavelength bands is mostly
decoupled from the ejecta matter (cf. Fig~\ref{fig:phmatrix}). The compositions
of the outer layers for the two base models are shown in Table~\ref{tab:comp}.
The models are calculated out to $v\sim70\,000\,\kms$ where the density is low
enough to ensure that the outer cutoff velocity does not affect the model
results. As for the original models, we use $500\,000$ Monte Carlo packets to
keep the Monte Carlo-induced noise in the UV spectrum sufficiently low.

\begin{table}
  \centering
  \caption{Outer-layer composition (mass fractions) of the two base models.
  Fe listed here corresponds to Fe not generated in the {\nifs}-decay.
  The decay products of {\nifs} after $t=29\,$d are listed separately below.}
  \begin{tabular}{lcc}
       & SN~2001eh & SN~2001ep \\
          \hline
          \hline
 $v_{\rm shell}$  & 15\,000\,\kms  &  14\,500\,\kms  \\
 C       & $2.0\times10^{-1}$&$8.0\times10^{-2}$\\
 O       & $7.7\times10^{-1}$&$8.3\times10^{-1}$\\
 Mg      & $1.0\times10^{-4}$&$1.0\times10^{-4}$\\
 Si      & $5.0\times10^{-3}$&$8.6\times10^{-2}$\\
 S       & $2.0\times10^{-2}$&$1.1\times10^{-3}$\\
 Ca      & $1.5\times10^{-3}$&$5.1\times10^{-4}$\\
 Ti      & $3.0\times10^{-4}$&$6.1\times10^{-5}$\\
 Cr      & $3.0\times10^{-4}$&$1.0\times10^{-5}$\\
 Fe      & $5.0\times10^{-4}$&$1.0\times10^{-3}$\\
 {\nifs} & $5.0\times10^{-3}$&$2.0\times10^{-3}$\\
 \hline
 {\nifs}      & $1.8\times10^{-4}$& $7.4\times10^{-5}$\\
 {\cofs}      & $4.0\times10^{-3}$& $1.6\times10^{-3}$\\
 {\fefs}      & $8.3\times10^{-4}$& $3.3\times10^{-4}$\\
  \hline
  \end{tabular}
  \label{tab:comp}
\end{table}

We chose to vary the composition of (i) Fe alone (corresponding to stable Fe
not produced via the {\nifs}-decay chain), (ii) {\nifs}  (at the epoch of $29\,$d
this corresponds to $3.7\%$Ni $79.7\%$ Co and $16.6\%$ Fe\footnote{using half
lives from \citet{junde99}}), and (iii) Ti and Cr together. The last two
elements do not produce strong visible features in the spectrum, but affect the
UV flux because they have a large number of relatively weak lines that densely
cover the UV and therefore contribute significantly to the total opacity.  For
each base model we increased and decreased the composition of the groups of
elements in steps of $0.5\,$dex.

\subsubsection{Results}

\begin{figure*}
  \begin{center}
    \includegraphics[width=8.3cm]{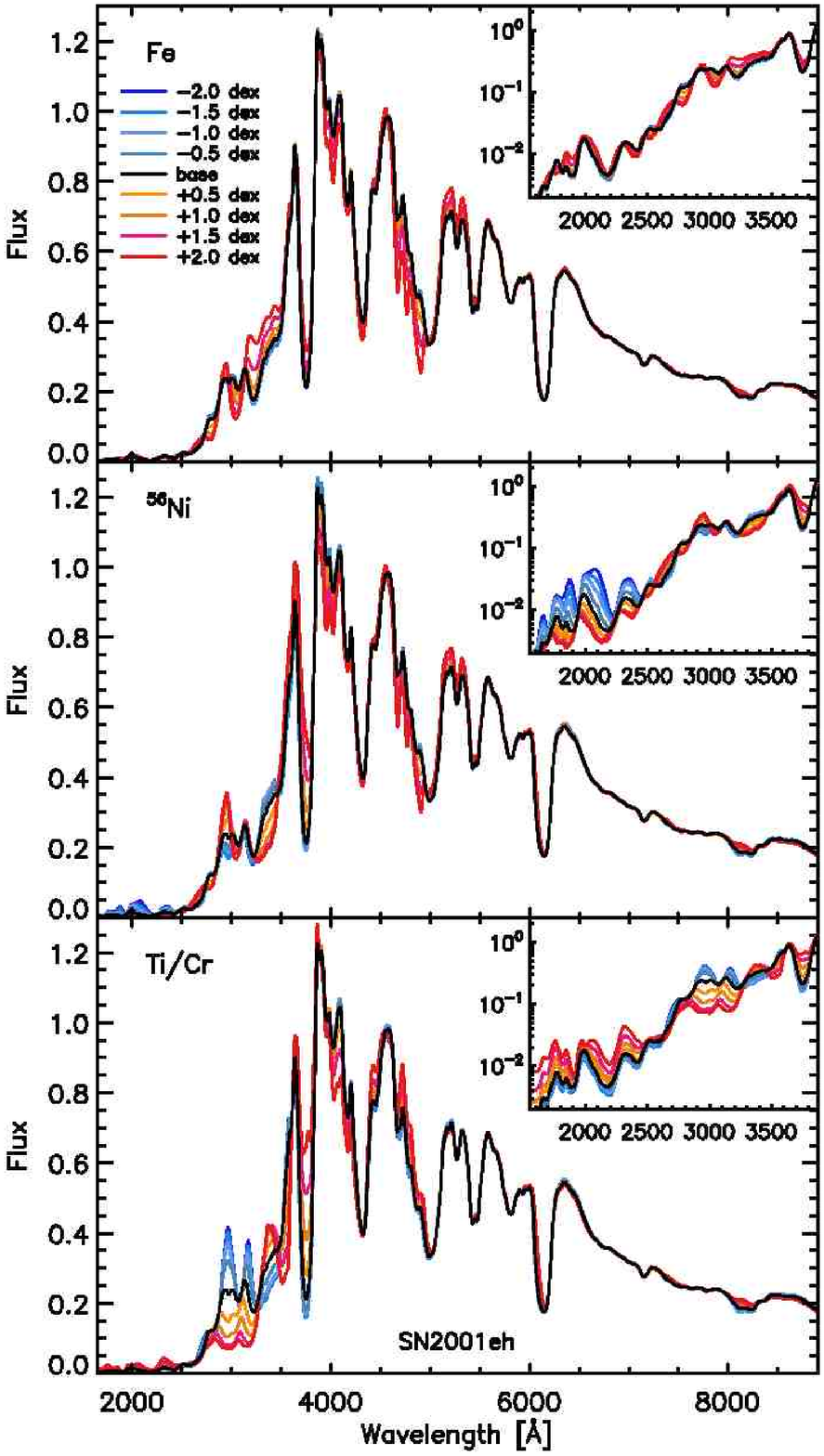}
    \includegraphics[width=8.3cm]{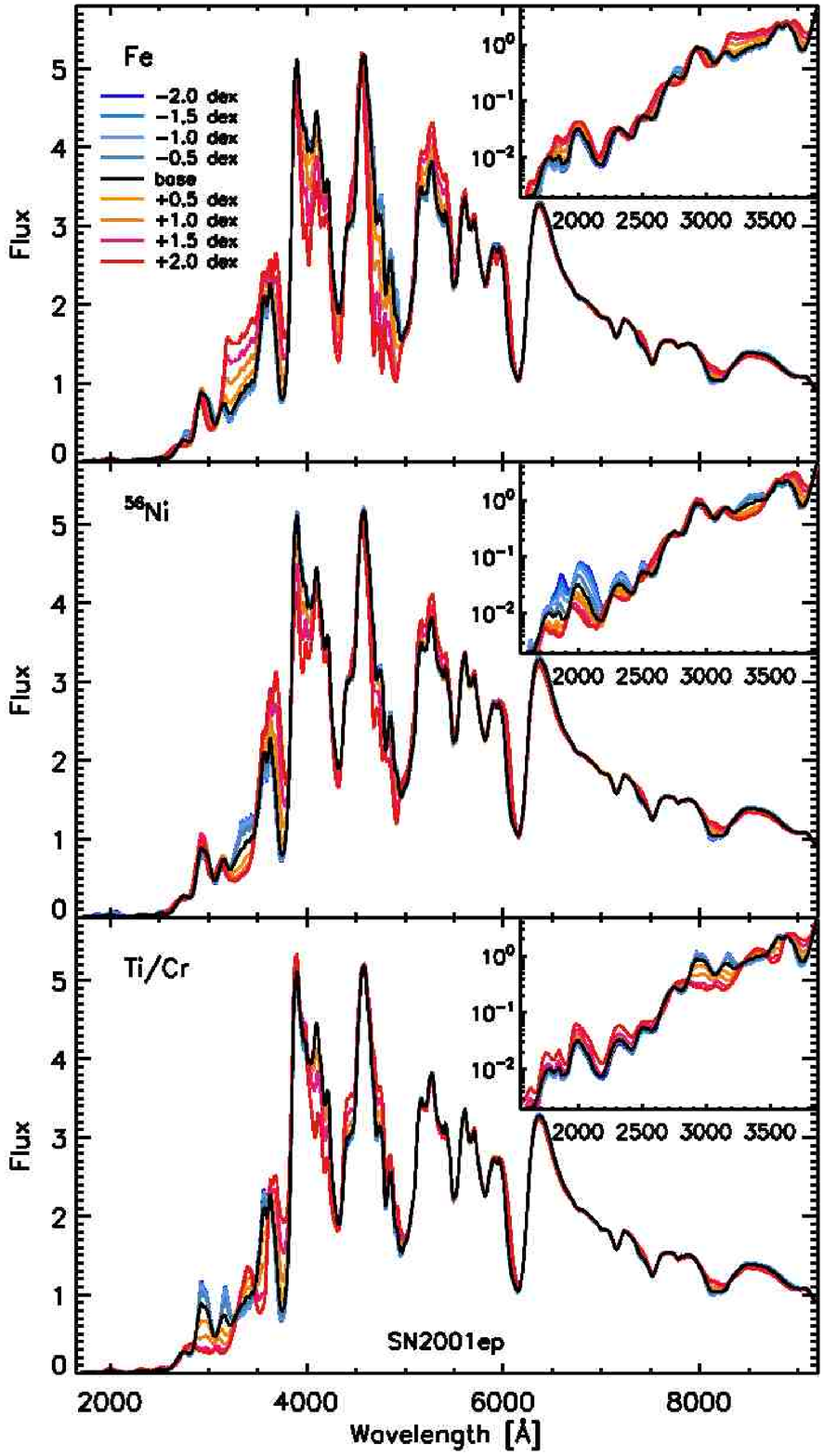}
  \end{center}
  \caption{Model series for varying composition of the outer layers for the
  SN~2001eh-based model series (left column) and the SN~2001ep-based model
  series (right column). The upper panels show the series where Fe alone is
  varied, the middle panels show the series with a variation of {\nifs} (i.e.
  Ni, Co, and Fe together), and the lower panels show the series for Ti and Cr.
  Colour coding indicates different models --- blue lines represent models
  where the abundance of the respective element is decreased, red lines show
  models where the abundance is increased.  The black line refers to the
  original model.} \label{fig:UvSeries}
\end{figure*}

The spectra of the series are shown in Fig.~\ref{fig:UvSeries}. The models in
the left column are based on SN~2001eh, the ones in the right column on
SN~2001ep. Each column shows the model for Fe variation alone in the upper
plot, the {\nifs} series in the middle, and the Ti/Cr series in the bottom
plot. The insets show the UV part of the spectrum in logarithmic scaling for
each set. The colour coding represents the different models -- the shades of
blue represent models in which the abundance of the element groups have been
decreased, while the shades of red are models in which the abundances have been
increased. For comparison the original model is shown in black.

Overall, an increase of abundances has a stronger effect than a decrease. This
is not surprising, because the abundance of heavy elements was already
relatively low in the outer layers of the base models. The comparison of the
three groups of models shows that the relative change of the optical part of
the spectrum is small to moderate; the variation of Fe alone has the strongest
effect on the optical part especially in the series of spectra based on
SN~2001ep (Fig.~\ref{fig:UvSeries}, right panels).

The variation of the blue and the UV part of the spectra differs among the
different groups of models.  In the SN~2001eh series the absorption feature at
$3050\,${\AA} becomes deeper with  higher Fe abundance; in the SN~2001ep series
this feature does not change significantly. The region red of this feature, up
to \tsim$3500\,${\AA}, shows significantly enhanced flux for higher Fe
abundance in both cases. In the SN~2001ep series this higher flux level also
affects the re-emission peak bluewards of the \ion{Ca}{2}~H\&K feature, which
remains mostly unchanged in the SN~2001eh models.  The models where Fe is
decreased are very close to the base model in both series.

Further to the UV, bluewards of \tsim$2500\,${\AA}, the models based on
SN~2001eh (Fig.~\ref{fig:UvSeries}, upper-left panel) show little change with
the variation of the Fe abundance. In the corresponding SN~2001ep-series
(Fig.~\ref{fig:UvSeries}, upper-right panel), we observe the trend that models
with higher Fe abundance have more flux in the UV, although the effect here is
also small.

The variation of {\nifs} is visible in a similar region as in the Fe models,
but with the opposite trend. Models with more {\nifs} have a deeper absorption
at \tsim$3300\,${\AA}, which is mostly caused by stronger \ion{Co}{3} lines.
Consequently, the emission peak bluewards of the \ion{Ca}{2} absorption becomes
stronger in those models (this peak is largely induced by the P-Cygni
re-emission of those Co-lines). Again the variation in this peak is stronger in
the SN~2001ep-based models. In the UV part blue of about $2500\,${\AA} the
effect of decreasing the {\nifs} abundance leads to  more flux while increasing
the {\nifs} abundance leads to less flux.

The bottom panels of Fig.~\ref{fig:UvSeries} show the models where the
abundances of Ti and Cr have both been varied. The most prominent result of
this variation is the  change of the double-peaked re-emission feature centred
at $3000\,${\AA}.  For low Ti and Cr abundances in the outer layers this
double-peaked structure becomes more and more prominent, while it almost
entirely disappears for high mass fractions of Ti and Cr. This behaviour is
more pronounced in the spectra of the SN~2001eh-series. In both series  high Ti
and Cr abundances result in an additional absorption feature at $3050\,${\AA},
mostly caused by \ion{Ti}{2}.  Further to the UV the trend again reverses,
leading to  more flux for the models with higher Ti and Cr abundance.  Blue of
\tsim$2700\,${\AA}  reductions of Ti and Cr have only a small effect.  In the
optical spectrum Ti and Cr predominantly affect the blue wing of the deep
absorption around $4300\,${\AA}.

\begin{figure}
  \begin{center}
    \includegraphics[width=8.3cm]{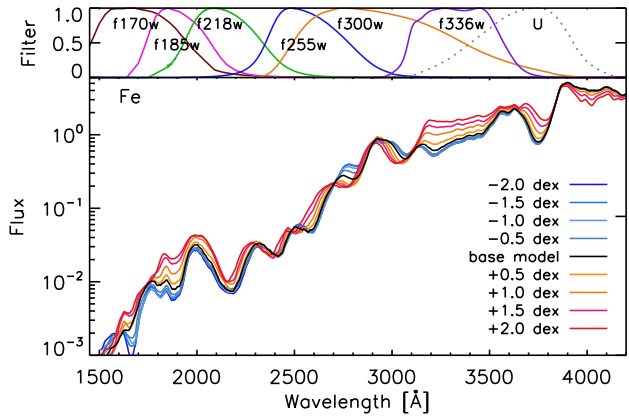}
  \end{center}
  \caption{Logarithmic plot of the SN~2001ep model series with varying Fe
  content (cf.  Fig.~\ref{fig:UvSeries}, upper-right panel). The upper panel
  shows the position and shape of the WFPC2 bandpass filters used for the
  comparison of integrated flux shown in Fig~\ref{fig:magsAbu01eh}.}
  \label{fig:epUvLog}
\end{figure}

To quantify the change in the models better we integrate the flux in different
bandpass filters of the WFPC2 instrument\footnote{See {\tt
http://www.stsci.edu/hst/wfpc2}} and the standard $U$-band filter
\citep{bessell90}. Fig.~\ref{fig:epUvLog} shows the logarithmic representation
of the Fe-model series of the SN~2001ep-based models in Fig.~\ref{fig:UvSeries}
(upper-right panel) together with the shape and position of the different
bandpass filters in the upper panel.

\begin{figure*}
  \begin{center}
    \includegraphics[width=8.3cm]{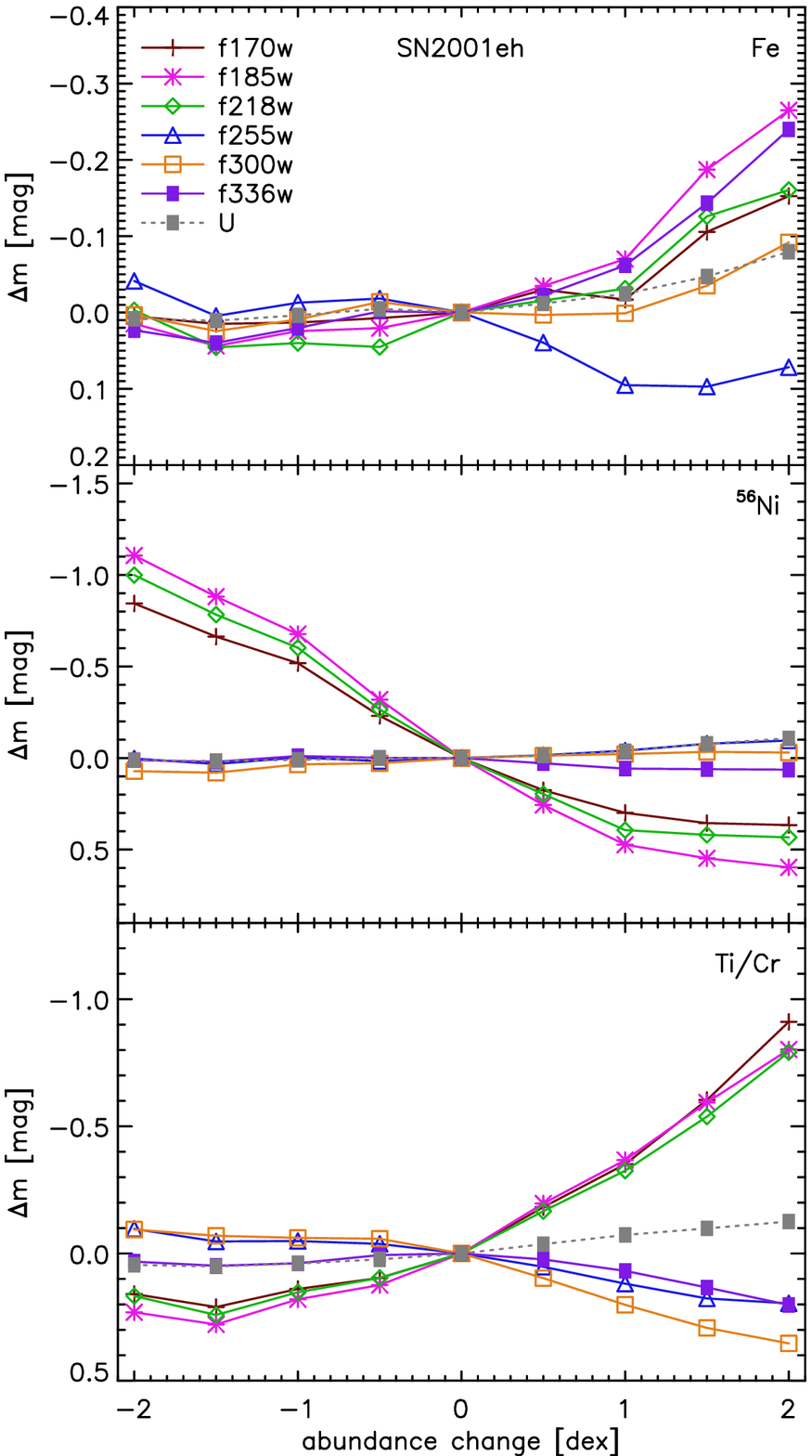}
    \includegraphics[width=8.3cm]{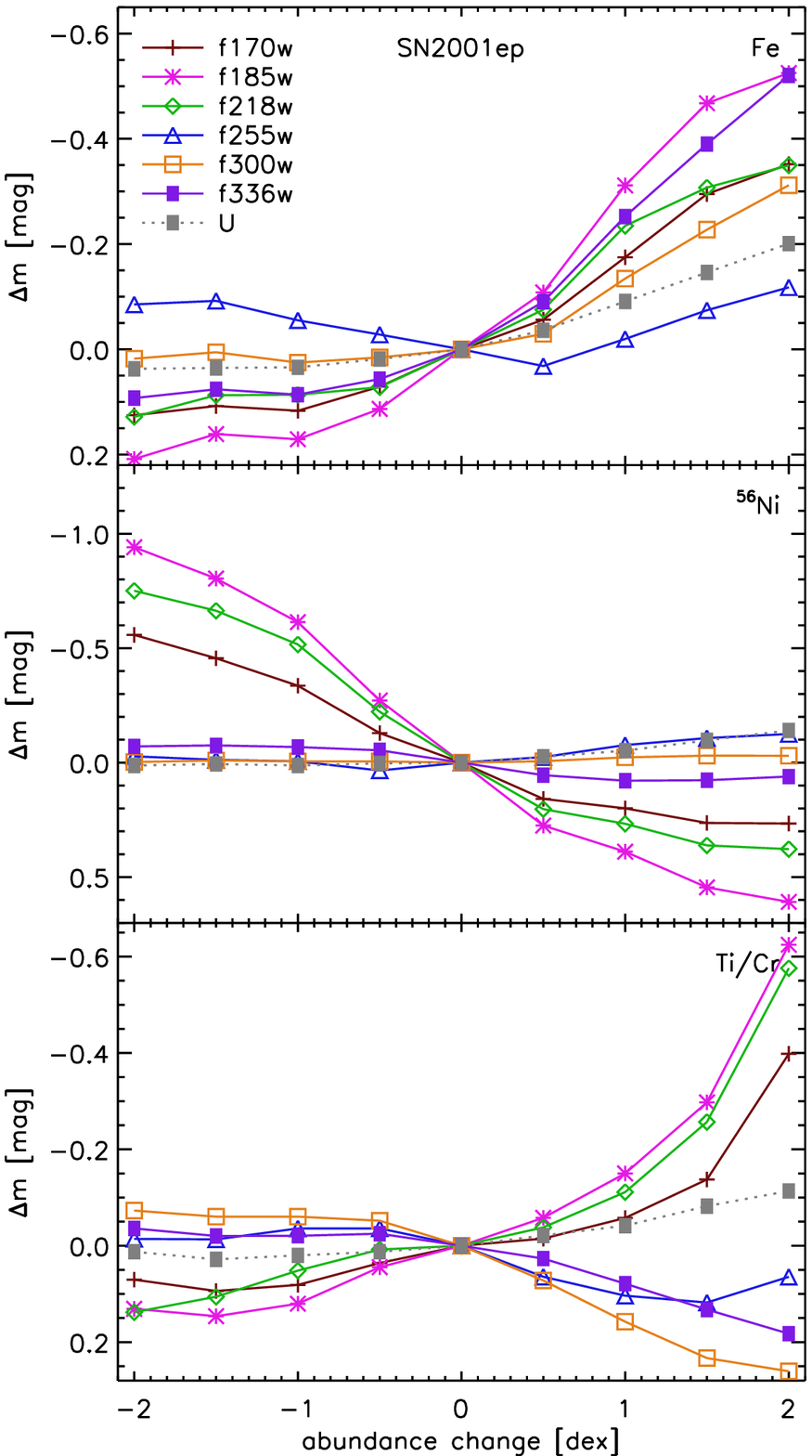}
  \end{center}
  \caption{Comparison of the integrated flux in different bandpass filters for
  the model series of SN~2001eh (left) and SN~2001ep (right).  The
  curves show the change in magnitudes in a specific filter band (cf.
  Fig.~\ref{fig:epUvLog}) relative to the base model versus the change in
  composition in the outer shell of the model.} \label{fig:magsAbu01eh}
\end{figure*}

Fig.~\ref{fig:magsAbu01eh}  shows the integrated flux of the model spectra as a
function of the change in the abundance of the respective set of elements
relative to the base model. The different lines and symbols correspond to the
filters indicated in the legend.  Smaller values of $\Delta m$ correspond to
more flux in the corresponding filter. Overall, the model series for the two
base models show similar behaviour, although the absolute numbers are slightly
different. The plots confirm what is visible in the spectra: the increase of Fe
and Ti/Cr leads to a higher flux in some passbands, while the modification of
{\nifs} has the opposite effect.  The $U$ band is almost unaffected by a
decrease of heavy-element abundances, while an increase always leads to more
flux (although to a small degree in some cases).  The feature that most
strongly affects the $U$-band is the re-emission peak bluewards of the
\ion{Ca}{2} H\&K absorption. The largest variation is usually seen in the
filter centred around $1850\,${\AA} (f185w).

\subsubsection{Discussion}

The change in the UV flux with varying composition is mediated by two main
effects. Firstly, an increased content of heavy elements with a large number of
lines in the UV increases the probability for the reverse fluorescence to
occur.  Secondly, the increased line opacity leads to a stronger line blocking
effect, which affects the temperature structure and consequently the ionization
structure in the ejecta.

\begin{figure}
  \begin{center}
    \includegraphics[width=8.3cm]{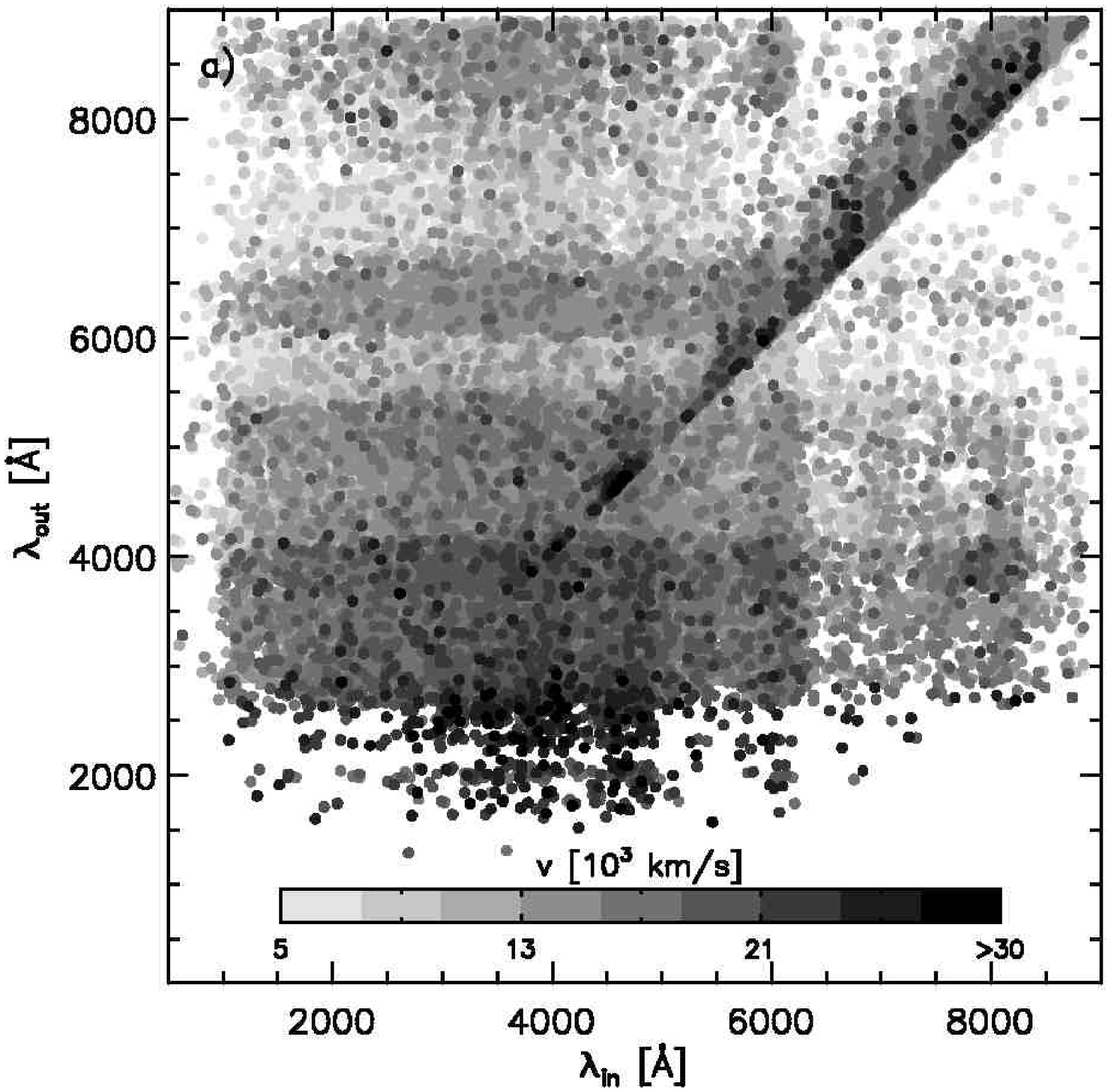}
    \includegraphics[width=8.3cm]{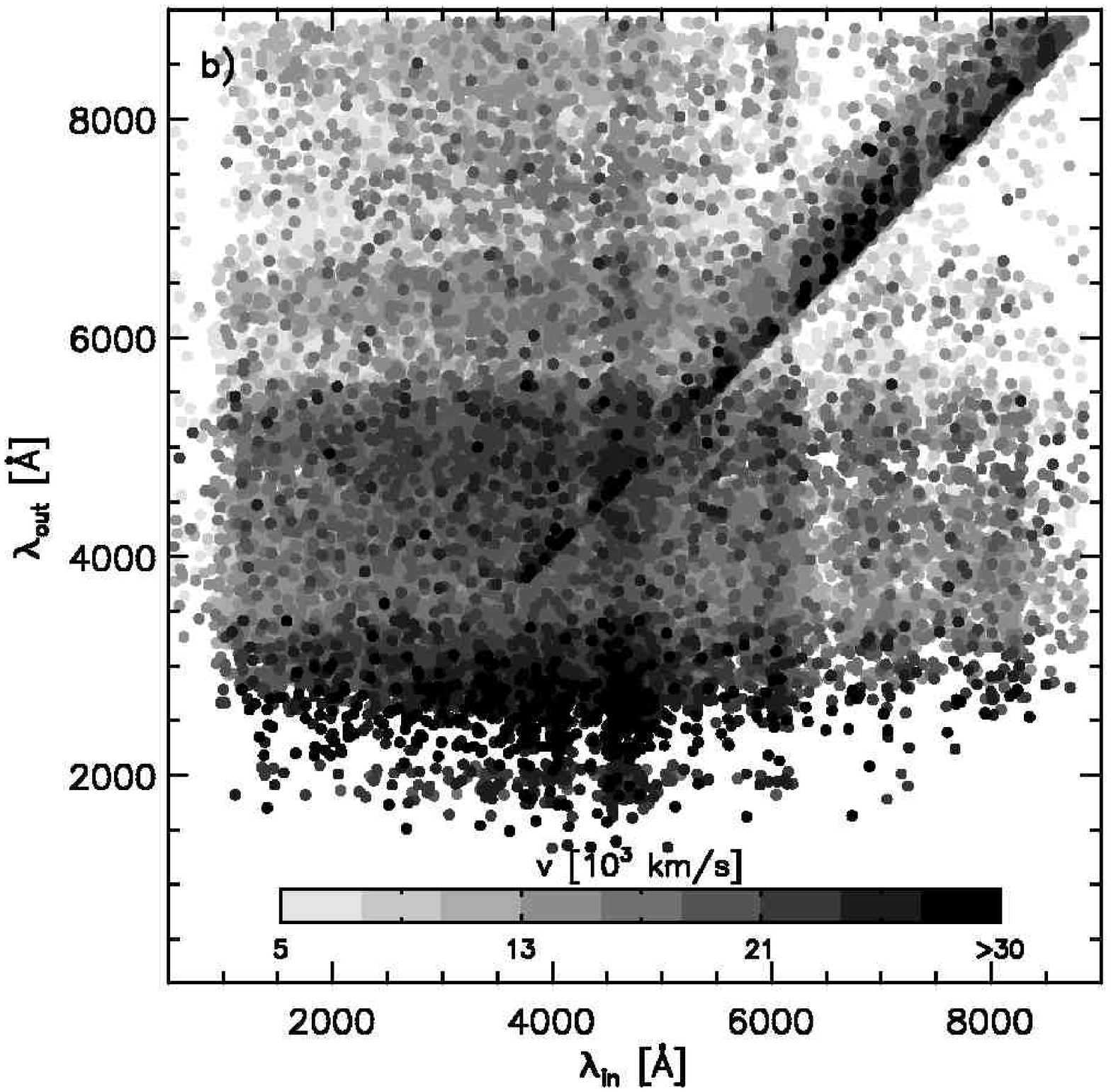}
  \end{center}
  \caption{Final versus initial wavelength of all photon packets that emerge to
  infinity \citep[cf.][]{mazzali00,pinto00b}.  The grey shading indicates the
  velocity of the last scattering event. The darker a point is, the higher the
  velocity of the shell at which the photon packet encountered the last
  interaction with a line.  (For wavelength pairs where more than one photon
  packet escapes, only the one with the highest velocity is shown. Note that a
  single point usually represents a larger number of photon packets.)  The
  diagonal line represents all photons that are resonance scattered.  All
  points above the diagonal are packets scattered to redder wavelengths, while
  all points below the diagonal represent packets that are scattered to bluer
  wavelengths through the reverse-fluorescence process.  The upper plot shows
  the base model of SN~2001ep; the lower plot shows the same quantities,
  however, for the model where the Fe abundance in the outer shell above
  $14\,500\,\kms$ was enhanced by $2\,$dex. The overall comparison shows that
  in the latter case, more photon interactions occur in the higher-velocity
  shells (more darker points). It also shows that more packets are scattered
  from an initially red wavelength to blue and UV wavelengths (dark points with
  outgoing wavelengths $\lesssim3500\,${\AA} below the diagonal).}
  \label{fig:phmatrix}
\end{figure}

The two plots in Fig.~\ref{fig:phmatrix} illustrate the reverse fluorescence
effect in the Monte Carlo radiative transfer model for SN~2001ep.  The points
indicate the initial wavelength at which a photon package has been created
versus the final wavelength at which it emerges from the envelope after the
last scattering event. The grey scale indicates the velocity of the shell where
the last interaction occurred. Darker points correspond to higher velocities.
Resonance scatterings that do not change the wavelength of the package are
located on the diagonal.

The data in the upper panel are from the simulation of the base model for
SN~2001ep; the data in the lower panel are from the model with high Fe
abundance in the outer shells. Both simulations have a similar total number of
emerging packets (\tsim$180\,000$), but the diagrams show only a fraction of
this total number because many packets are located on identical positions in
the wavelength grid.  In the model with higher Fe abundance in the outer layers
the photons encounter more line events and are therefore more likely to be
scattered out of their original wavelength.  In the base model $10\%$ of all
packets emerging with $\lam<3750\,${\AA} encounter their last interaction in a
shell with $v>20\,000\,${\kms}; in the model with high Fe abundance this
fraction is $20\%$.  Counting all packets emerging from the envelope regardless
of wavelength, the fractions of packets encountering the last interaction at
$v>20\,000\,${\kms} are $3.9\%$ (base model) and $5.6\%$ (Fe enhanced).

The condition for the reverse fluorescence process to lead to an increase of
the UV flux is that the UV lines are saturated such that photons can escape
from the lines without being re-absorbed immediately. Part of the UV
enhancement seen in the models with increased Fe abundance is therefore caused
by the presence of a larger number of saturated Fe lines in the UV.  The
decrease of UV flux in the models with enhanced {\nifs} abundance indicates
that the dominating Co lines in those models are on average less saturated.
Therefore the optical depth in those lines increases with abundance.

The second consequence of the abundance variation in the outer shells is a
change in the temperature and ionization structure caused by the change in line
opacity. Fe group elements have a large number of lines in the UV and blue part
of the spectrum. The high velocities cause them to overlap and effectively
prevent photons to emerge. The higher radiation energy density in the ejecta
leads to higher temperatures and consequently a higher state of ionization.

\begin{figure}
  \begin{center}
    \includegraphics[width=8.3cm]{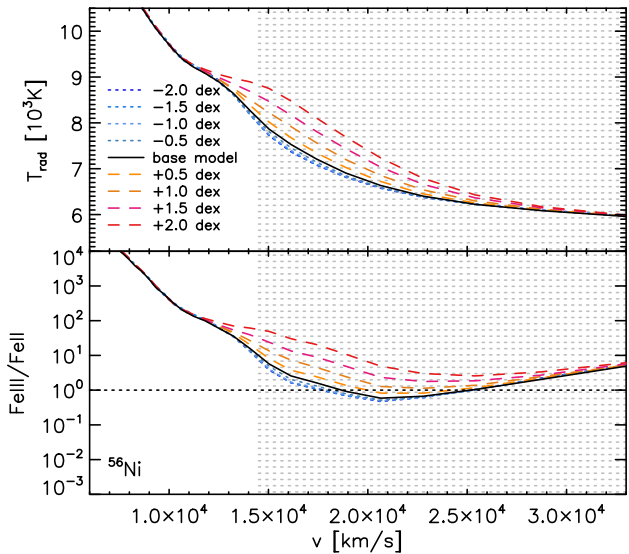}
  \end{center}
  \caption{The upper panel shows the temperature as a function of
  velocity in the SN~2001ep model series where the abundance of {\nifs} was
  varied in the outer region above $v_{\rm shell}=14\,500\,${\kms} (shaded
  area). The lower panel shows the relative ionization of
  \ion{Fe}{3}/\ion{Fe}{2} for the same models.}  The models differ slightly
  even below $v_{\rm shell}$ because of the different number of photons scattered
  back into the envelope from outer layers depending on the heavy-element
  abundance. \label{fig:ionf99}
\end{figure}

Fig.~\ref{fig:ionf99} shows the mean radiation temperature (upper panel) and
the ratio of doubly to singly ionized Fe  (lower panel) for the SN~2001ep-based
model series in which {\nifs} is varied above $14\,500\,${\kms} (shaded area).
A change to lower abundances does not have a strong effect but with increasing
{\nifs} abundance a clear shift to higher temperatures and consequently doubly
ionized species is induced. Backwarming extends the region affected to
velocities lower than $v_{\rm shell}$ above which the actual variation has been
imposed.

\begin{figure}
  \begin{center}
    \includegraphics[width=8.3cm]{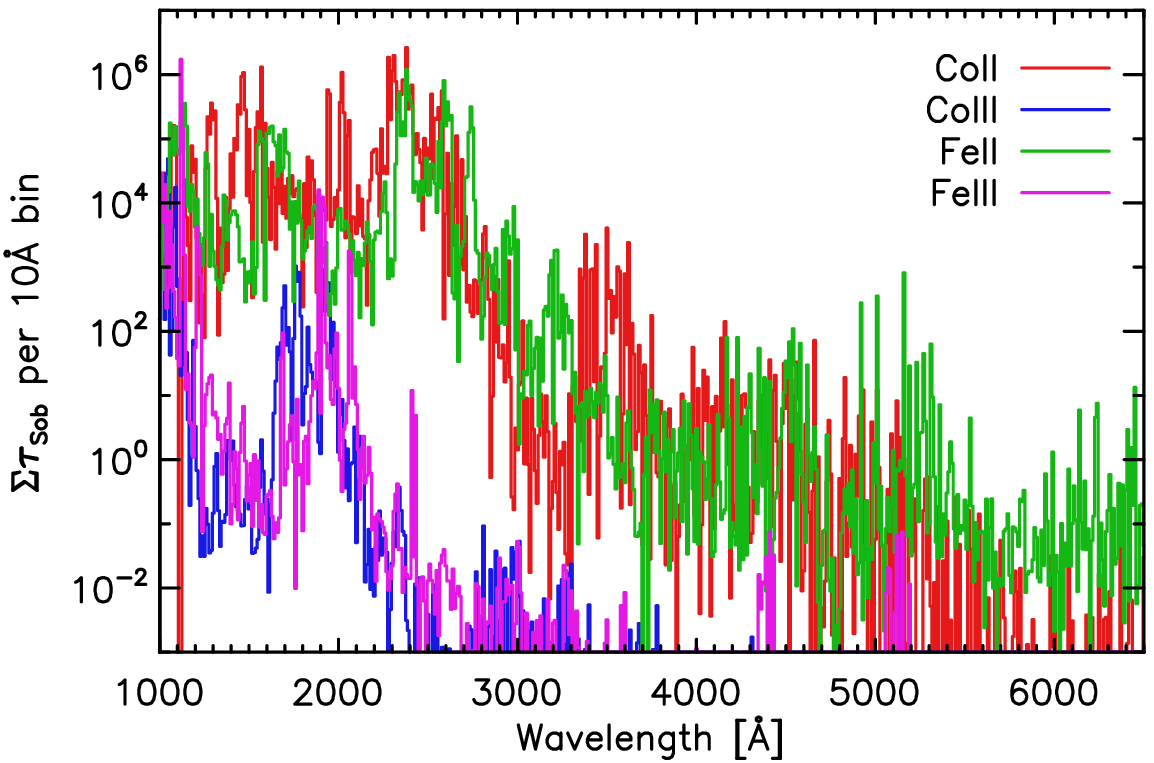}
    \includegraphics[width=8.3cm]{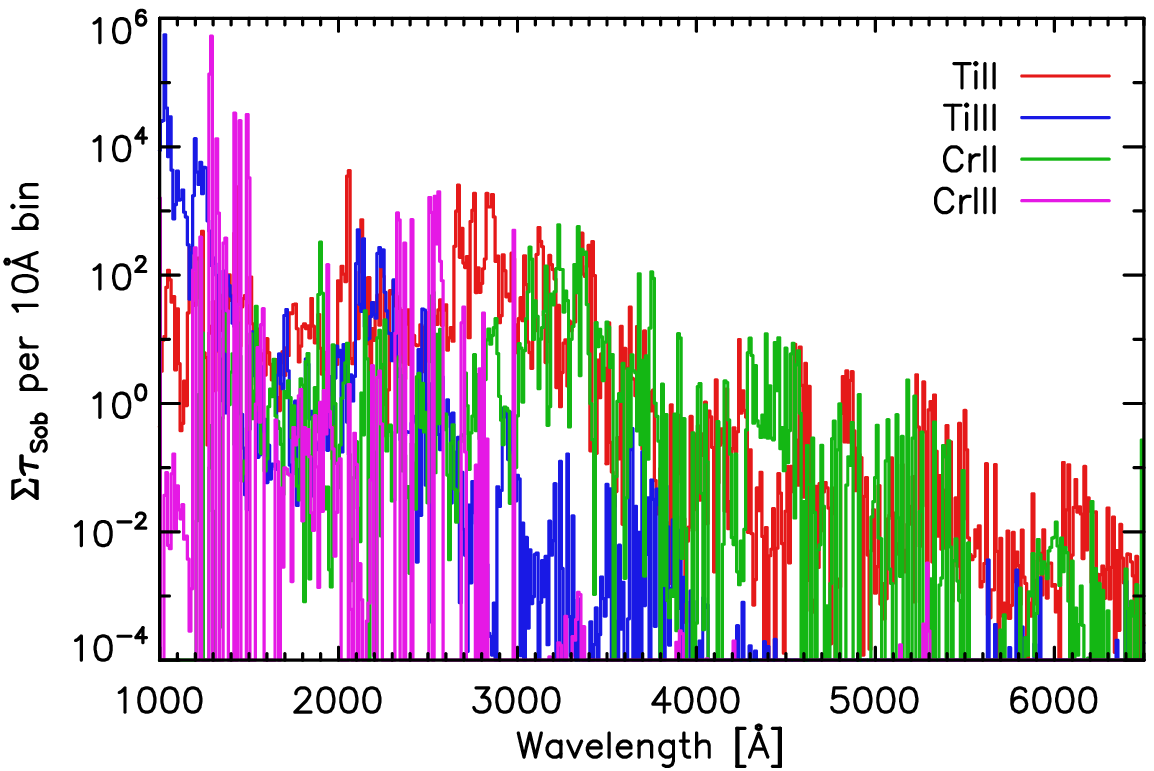}
  \end{center}
  \caption{Sum of $\tau_{\rm Sob}$ for all lines in a wavelength interval of
  $10\,${\AA} for the ions indicated in the plot legend. The Sobolev optical
  depths are derived for the model of SN~2001ep.}
  \label{fig:tlines}
\end{figure}

\citet{pinto00b} discuss that in a gas having the physical conditions typically
found in supernovae (low matter densities, high radiation energy densities) and
dominated by sufficiently complex ions with a large number of possible line
transitions, an equilibrium of the radiation field is established solely by
radiative transitions. They regard this as an additional thermalization process
that is not related to collisional and true continuum processes. In contrast to
real thermalization, however, this equilibrium state does not couple the
thermal properties of the gas to the radiation field.  Therefore, the emitted
spectrum depends only weakly on the temperature of the gas (as long as the
ionization does not change significantly with temperature).  Instead, the
spectrum reflects the distribution of line strengths as a function of
wavelength.  A change in the dominant species in the outer layers, either by a
variation of the composition or the ionization state of the gas,  can affect
the density of spectral lines in the UV  and may  have a strong effect on the
emergent spectrum. Fig.~\ref{fig:tlines} shows the distribution of line
strengths in terms of Sobolev optical depths for some of the relevant ions
found in the mixture of SN~Ia\footnote{The data shown in this plot are
extracted from the SN~2001ep model but the distributions depend only weakly on
the model.}.  The qualitative difference of the distribution between different
elements explains why different wavelength regions are affected when Ti and Cr
are varied compared to the variation of  Fe and Co.  Overall the higher
ionization stages tend to have fewer strong lines in the relevant spectral
region.  Therefore, a higher ionization decreases the total opacity.  This
introduces a feedback effect that prevents the opacity from growing arbitrarily
with the abundance of Fe-group elements.

\subsection{Variation of the density structure}

\subsubsection{Setup}

\begin{figure}
  \begin{center}
    \includegraphics[width=8.3cm]{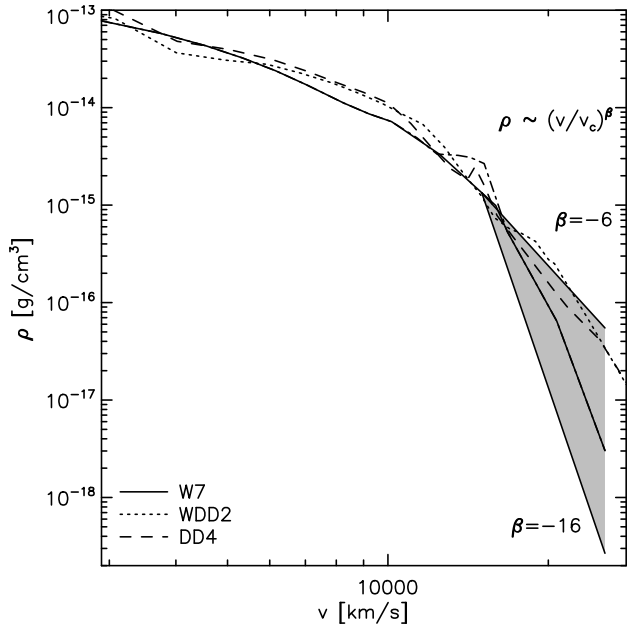}
  \end{center}
  \caption{The model setup for the variation of the density structure in the
  outer part. The W7 density (solid line) is replaced by a power-law with
  different power-law indices $\beta$ between $-5$ and $-16$ (shaded area). For
  comparison the two delayed-detonation models WDD2 \citep{iwamoto99} and DD4
  \citep{woosley94a} are also shown.  For the spectral models the density
  structures are extrapolated to higher velocities (\tsim$70\,000\,${\kms}) to
  avoid artefacts resulting from the outer boundary.} \label{fig:dens}
\end{figure}

For this series we kept the inner structure of W7 out to a velocity
$v_{\rm shell}=15\,000\,${\kms} and replaced the outer part with a power-law
density $\rho \propto (v/v_{\rm shell})^{\beta}$ with power-law index $\beta$
ranging from $-5$ to $-16$. The grey shaded area in Fig.~\ref{fig:dens} shows
the range of variation considered. For comparison, the density structures of
the pure 1D deflagration model W7 \citep{nomoto84} and the delayed detonation
explosion models DD4 \citep{woosley94a} and WDD2 \citep{iwamoto99} are also
plotted. The DDT models have somewhat more mass at high velocities (i.e. a
shallower density gradient). The variation considered in the parametrized
models covers the typical range of gradients for deflagration and delayed
detonation models.  All models are extrapolated to $v\sim70\,000\,${\kms}
assuming power-law densities to ensure that the outer boundary does not affect
the results.

Results the from multi-dimensional explosion models of \citet{roepke07b}
indicate that the ``bump'' and the relatively sharp change in the gradient of
the W7 deflagration model around $13\,700${\kms} may be an artefact of the
one-dimensional treatment.  Multi-dimensional models tend to have a smoother
density structure, although this can depend strongly on the assumptions made
when averaging multi-dimensional models over angles to obtain a 1D density
structure. If the deflagration phase is followed by a delayed detonation,
material is burned to higher velocities than in a pure deflagration because the
flame is accelerated and can reach the outer regions of the white dwarf before
the densities are too low for burning to proceed \citep{roepke07}.

At the epoch considered here, a modification of the density gradient at high
velocities is expected to affect the UV part of the spectrum most strongly
because the region where the optical spectrum is formed lies at much lower
velocities.  We restrict this study to one-dimensional explosion models because
our radiative transfer models are spherically symmetric.

\subsubsection{Results}

\begin{figure}
  \begin{center}
    \includegraphics[width=8.3cm]{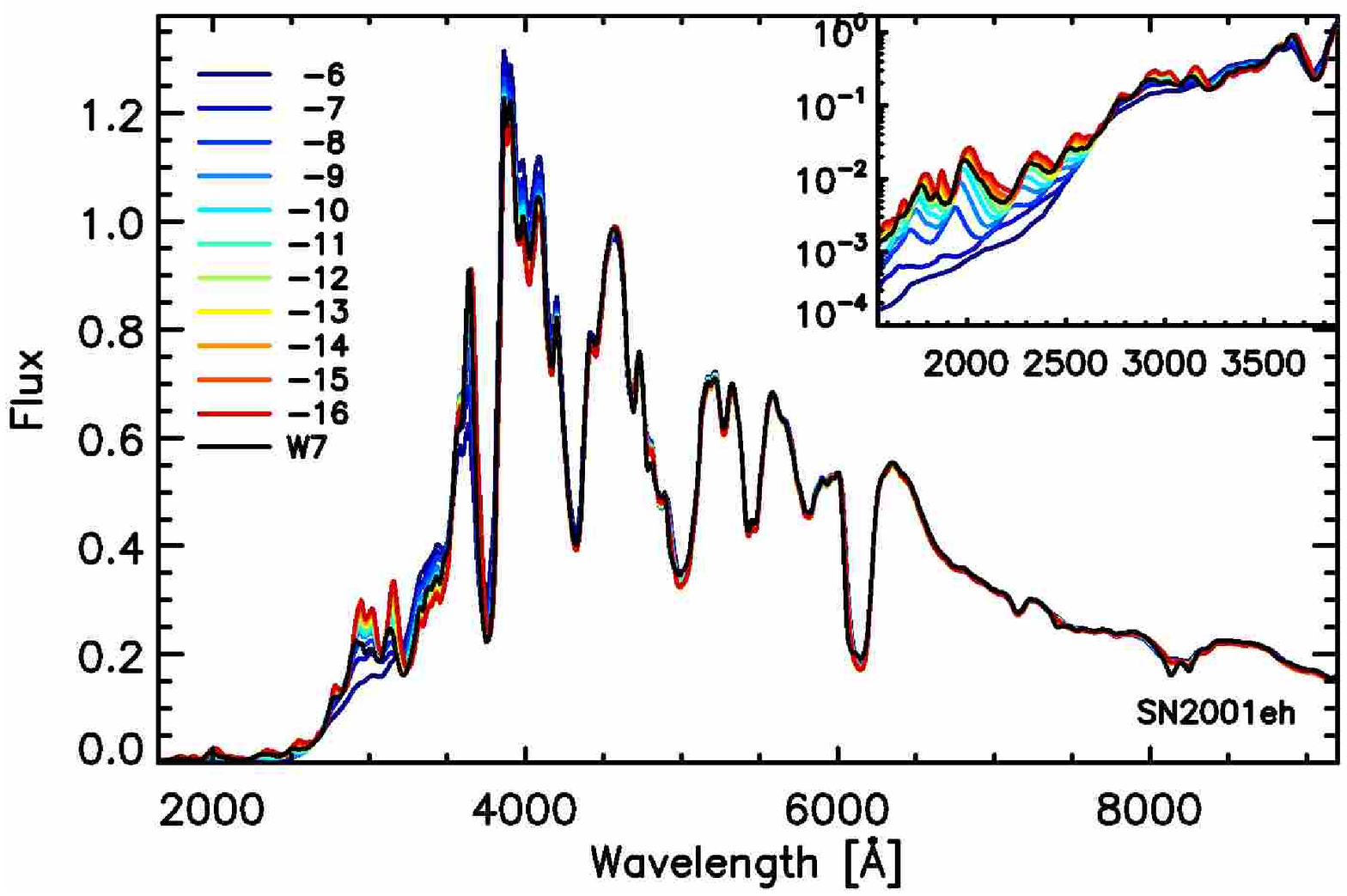}
    \includegraphics[width=8.3cm]{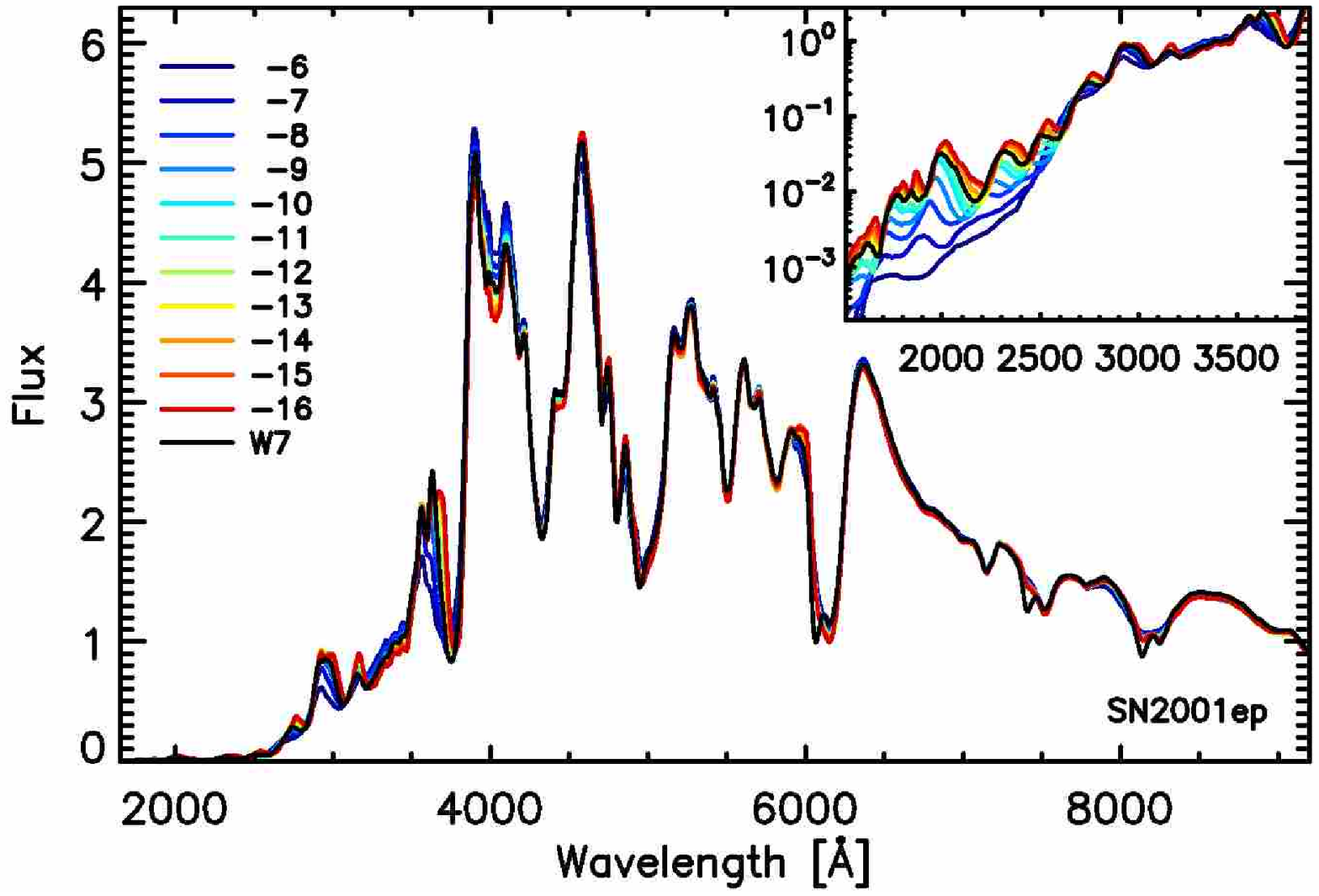}
  \end{center}
  \caption{The full UV and optical spectra of the model series with varying
  power-law density above $v_{\rm shell}=15\,000\,${\kms}. The upper panel
  shows the models based on the model for SN~2001eh, the lower panel shows
  those based on SN~2001ep. The colour coding refers to the power-law index
  $\beta$ used for each model.  For comparison, the black line refers to the
  original W7-based model spectrum, which is close to the $\beta=-12$ case in
  the outer part. The insert shows the UV part of the spectrum in logarithmic
  scaling.} \label{fig:varDens}
\end{figure}

Fig.~\ref{fig:varDens} shows the model series for different gradients as well
as the model based on the W7 density structure.  The region which is mostly
affected by the change is the UV bluewards of about $4200\,${\AA}. The rest of
the spectrum remains largely unchanged.

The W7 model corresponds roughly to the model with $\beta=-12$. Models with a
comparable or steeper gradient than the W7 structure show similar features that
tend to become narrower for steeper gradients.  In the models with shallower
gradients the features tend to become broader and shift bluewards for a
power-law index of $\beta \lesssim -9$.  In even shallower models the
structures become progressively less distinct and the flux in the UV below
$2750${\AA} drops significantly. The overall effect is similar in both model
series except for the double peaked emission feature centred around $3000\,${\AA},
which shows a stronger variation in the series of SN~2001eh.

\begin{figure}
 \begin{center}
   \includegraphics[width=8.3cm]{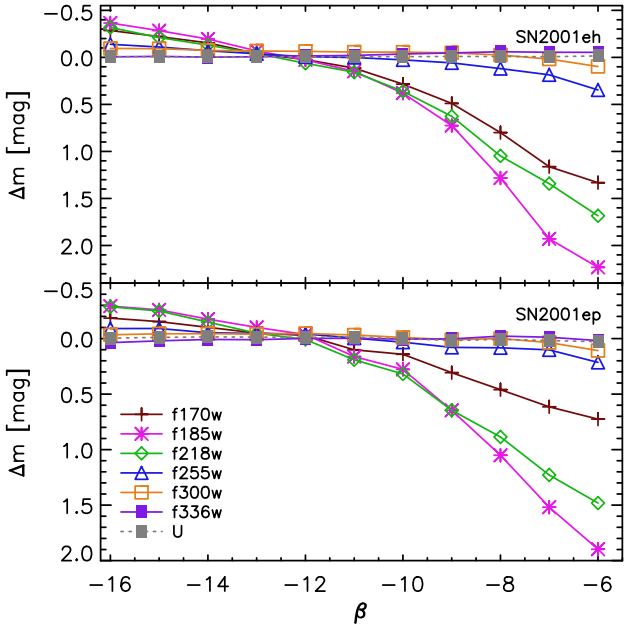}
 \end{center}
   \caption{Variation of the integrated flux in the filters shown in
   Fig.~\ref{fig:varDens} relative to the original W7-based models.}
   \label{fig:densFilters}
\end{figure}

Fig.~\ref{fig:densFilters} shows the integrated flux in the bandpass filters
shown in Fig.~\ref{fig:epUvLog} for the model series with density variation
(Fig.~\ref{fig:varDens}).  The strongest variation is observed for the three
bluest filters (f170w, f185w, and f218w), while other bands show only minor
changes.  For very steep gradients the flux is enhanced in those bands relative
to the base model. As the gradient flattens the flux drops by up to $2.4$mag in
the f185w-band.  As expected, the results  for the two supernovae are very
similar.  The difference at \tsim$3000\,${\AA} is not visible in the integrated
flux because the f300w filter is too broad.

\begin{figure}
  \begin{center}
    \includegraphics[width=8.3cm]{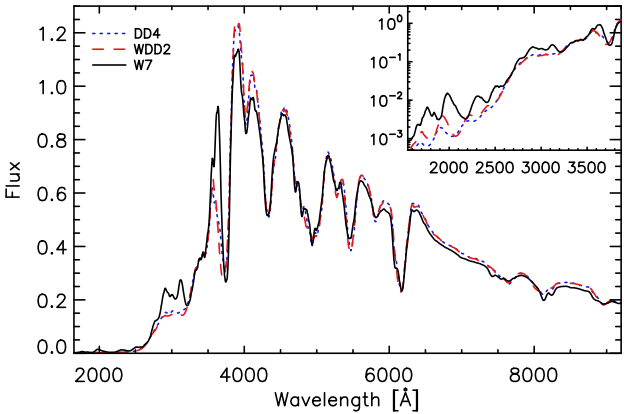}
  \end{center}
  \caption{Spectra of the one-dimensional delayed detonation explosion models
  WDD2 and DD4 as well as the deflagration model W7 assuming an identical
  composition for all models. The series shown is based on the model for SN~2001eh.}
  \label{fig:specExplModel}
\end{figure}

We also computed models for the density structures of  the two one-dimensional
delayed detonation models WDD2 \citep{iwamoto98} and DD4 \citep{woosley94a}
using the same composition as in the base model for SN~2001eh. The resulting
spectra are shown in Fig.~\ref{fig:specExplModel}.  Both of the delayed
detonation explosion models have a somewhat higher kinetic energy than W7 and
thus more mass at higher velocities (cf.  Fig.~\ref{fig:dens}). The density
gradients in the outer parts of both models are comparable and, therefore, the
model spectra are not very different.  Compared to the W7 model, the prominent
features in the blue and UV are broader and shifted bluewards in the delayed
detonation models. The same comparison for SN~2001ep shows qualitatively
similar features but the difference to W7 is smaller.

\subsubsection{Discussion}

\begin{figure}
  \begin{center}
    \includegraphics[width=8.3cm]{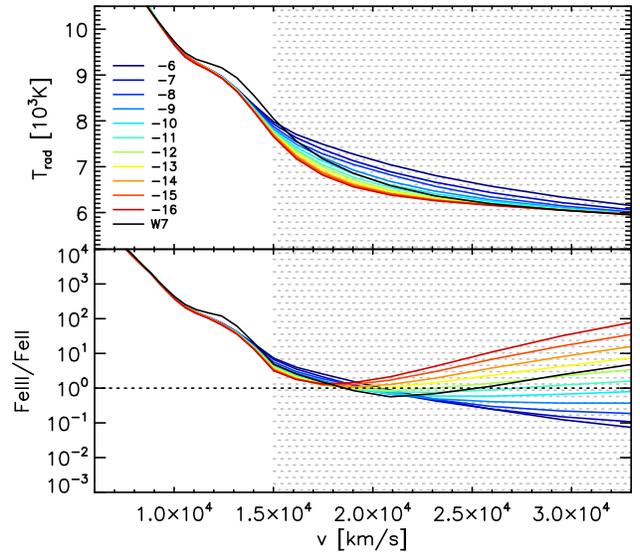}
  \end{center}
  \caption{Similar plot as in Fig.~\ref{fig:ionf99} for the models with varying
  density gradients based on the model of SN~2001ep. The colour coding of the
  models is the same as in Fig.~\ref{fig:varDens}. The W7-model roughly
  corresponds to $\beta=-12$, however is somewhat shallower at lower
  velocities. } \label{fig:IonfDenseh}
\end{figure}

Line opacities scale with density. Therefore the modification of the density
gradient in the outer layers affects the wavelength range over which a photon
can interact with a spectral line. Shallower density structures increase the
effective range of Doppler-shifts from the line-centre letting more lines
overlap and leading to broader line features.  This makes the  line blocking at
UV wavelengths even more efficient.  Fig.~\ref{fig:IonfDenseh} shows the
temperatures and the ionization ratio \ion{Fe}{3}/\ion{Fe}{2} for the
SN~2001ep-model series with a varying density gradient.  While the temperature
varies only mildly for models with a density gradient steeper than W7, it
becomes much higher in the outer parts for the models with very shallow density
gradients.  As in the models with varying composition this behaviour reflects
the increased heating of the ejecta by the higher radiation density caused by
the more efficient line blocking.  The ionization pattern shown in the lower
panel of Fig.~\ref{fig:IonfDenseh} roughly follows the temperature, with the
exception that models with very steep gradients show a significant shift to
higher ionization stages in the outer parts beyond $v\approx17\,000\,${\kms}.
Here the low densities become the governing factor that drives the ionization
towards the doubly ionized species although the densities are too low to leave
a significant imprint in the absorption features in the optical spectrum.  The
pattern is very similar for the other elements.  The role of the distribution
of line strength discussed in the previous subsection applies to this model
series in the same way. The varying spectral appearance in the UV can therefore
be qualitatively understood by the shift in ionization for the elements that
provide the major source of opacity.

\section{Conclusions}
\label{sec:conclusions}

We analysed the spectra of two {\sneia} that have spectral coverage in the UV
wavelength bands. Although the spectra were taken at very similar epochs, their
appearances are different. The models show that the typical temperatures in the
ejecta of SN~2001eh are somewhat higher than in SN~2001ep. The shape of the
spectrum of SN~2001eh is determined by absorptions of doubly ionized species
(\ion{Fe}{3} and \ion{Co}{3}) while the shape of the spectrum of SN~2001ep is
determined by lines of singly ionized iron-group elements (\ion{Fe}{2} and
\ion{Co}{2}).  The difference is especially visible in the UV bluewards of
\ion{Ca}{2} H\&K.  Based on these spectral models we investigated the
dependence of the UV flux on the metallicity and the density gradient in the
outer layers of the ejecta. In one series of models we varied the abundance of
groups of heavy elements; in another series we modified the density gradient in
the outer layers and replaced the adopted explosion model with a power-law
density structure that has different power-law indices.

The model series with varying composition shows that under certain conditions
the UV flux may actually {\em increase} with increasing abundance of heavy
elements while other models show the opposite behaviour. The physical process
that dominates the formation of the UV flux in {\sneia} is a
reverse-fluorescence process in which photons are absorbed at long wavelengths
and re-emitted at a shorter wavelength in the blue or UV. Therefore, the shape
of the spectrum is determined by the distribution of spectral lines as a
function of wavelength.  This distribution varies slightly for different
elements, but is very sensitive to different ionization stages. A temperature
change that results in a shift of the ionization balance directly affects the
efficiency of line blocking in the UV, which couples back to the temperature.
This feedback results in a non-linear response of the UV flux to changes in the
metallicity.  This wavelength region is therefore a valuable probe for the
outer regions of explosion models, which do not leave a strong signature at
longer wavelengths. To further study this effect quantitatively, a more
realistic treatment of the temperature structure, consistent with the non-LTE
state of the gas, should be used. In this work we can only give a qualitative
assessment of potential variations.

The density gradient mostly affects the breadth of observed line features.
Shallow density gradients increase the velocity region over which a line can
efficiently absorb and re-emit photons. For very shallow density structures the
densely clustered lines from Fe group elements in the UV cause individual
absorption features to merge together  resulting in a strong depletion of the
flux. Steeper density gradients lead to narrower and more pronounced absorption
and re-emission features.

Overall, the model series of both supernovae show the same qualitative changes.
The main differences seen in the UV affect the shape and strength of the
re-emission peak at \tsim$3500\,${\AA}, blue of \ion{Ca}{2} H\&K.

To conclude, a variation in composition and density of the high-velocity layers
of the ejecta, which may be introduced by different progenitor environments or
different degrees of mixing in the actual explosion, can have a significant
impact on the observed UV flux even at later epochs when the optical spectrum
is formed at much deeper layers.  Depending on the exact physical conditions, a
variation of metallicity can either reduce or increase the UV flux without
leaving a strong imprint on the spectrum at longer wavelengths. For a better
estimate of the magnitude and the scatter of the variation found in real
supernovae, more UV observations of {\sneia}, preferably over a range of
epochs, are needed. With the upcoming repair of STIS and the installation of
the Cosmic Origins Spectrograph (COS) on {\it HST}, it should be possible to
obtain additional UV spectra of {\sneia} that would be critical for such studies.

\section*{acknowledgements}

We thank Stuart Sim for many helpful discussions and Stefan Taubenberger for
his help sorting out the WFPC2 filter functions. We also thank the anonymous
referee for many constructive suggestions which helped to improve the
manuscript. D.N.S.  acknowledges support from the European Union's Human
Potential Programme {\em ``Gamma-Ray Bursts: An Enigma and a Tool,''} under
contract HPRN-CT-2002-00294 and the Transregional Collaborative Research Centre
TRR33 {\em ``The Dark Universe''} of the Deutsche Forschungsgemeinschaft.  This
research is supported by NASA/{\it HST} grants GO--9114 and GO--10182 from the
Space Telescope Science Institute, which is operated by the Association of
Universities for Research in Astronomy, Inc., under NASA contract NAS 5-26555.
A.V.F. is grateful for financial assistance from the National Science
Foundation (NSF grant AST--0607485), the Sylvia and Jim Katzman Foundation, and
the TABASGO Foundation, without which the construction and continued operation
of KAIT would have been impossible.  Support for supernova research at Harvard
University is provided in part by NSF grant AST--0606772. D.N.S., P.A.M., S.B.,
and R.P.K. thank the Kavli Institute for Theoretical Physics, at the University
of California, Santa Barbara, for its hospitality during the program {\em
``Accretion and Explosion: the Astrophysics of Degenerate Stars,''} supported
in part by the NSF under grant PHY--0551164. This research has made use of the
NASA/IPAC Extragalactic Database (NED), which is operated by the Jet Propulsion
Laboratory, California Institute of Technology, under contract with NASA.

\end{document}